\PassOptionsToPackage{dvipsnames,table}{xcolor}
\documentclass[a4paper,11pt]{article}

\usepackage{geometry}
\usepackage{xcolor}
\usepackage{microtype}
\usepackage{eso-pic}
\usepackage{amsmath, amssymb}
\usepackage{newtxtext,newtxmath}
\usepackage[]{graphicx}
\usepackage{sectsty}
\allsectionsfont{\sffamily}
\usepackage{fancyhdr}
\PassOptionsToPackage{hyphens}{url}
\usepackage{hyperref}
\usepackage{doi}
\usepackage[round]{natbib}
\usepackage{orcidlink}
\usepackage{booktabs}
\usepackage{multirow}
\usepackage{alltt}
\usepackage[labelfont={bf,sf},font=small]{caption}
\usepackage[title]{appendix}
\usepackage{nameref}
\usepackage[onehalfspacing]{setspace}
\usepackage[labelfont=bf,font=small]{caption}

\makeatletter
\def\maxwidth{ %
  \ifdim\Gin@nat@width>\linewidth
    \linewidth
  \else
    \Gin@nat@width
  \fi
}
\makeatother

\usepackage{framed}
\makeatletter
 {\par\unskip\endMakeFramed%
 \at@end@of@kframe}
\makeatother

\definecolor{unipd_red}{HTML}{b40908}

\newcommand\mail{roberto.macridemartino@deams.units.it}
\title{\vspace{-2em}
  \textbf{\textbf{Mixture priors for replication studies}}
}
\author{
   \textbf{Roberto Macrì-Demartino}\thanks{Corresponding author e-mail: \href{mailto:\mail}{\texttt{\mail}}} \textsuperscript{ $a$}   \orcidlink{0000-0002-5296-6566}, \textbf{Leonardo Egidi}\textsuperscript{$a$} \orcidlink{0000-0003-3211-905X}, \textbf{Leonhard Held}\textsuperscript{$b$}  \orcidlink{0000-0002-8686-5325}, and \textbf{Samuel Pawel}\textsuperscript{$b$}  \orcidlink{0000-0003-2779-320X}  \\
  \small\textsuperscript{$a$} Department of Economics, Business, Mathematics, and Statistics ``Bruno de Finetti", University of Trieste,\\
  \small Via A. Valerio 4/1, Trieste, 34127, Italy\\
  \small\textsuperscript{$b$} Epidemiology, Biostatistics and Prevention Institute, Center for Reproducible Science,
	University of Zurich,\\
  \small Hirschengraben 84, Zurich, 8001, Switzerland \vspace{2em} \\
  {\color{unipd_red}\small {}}
}
\date{}

\IfFileExists{upquote.sty}{\usepackage{upquote}}{}

\usepackage{hyperref}
\hypersetup{
  bookmarksopen=true,
  breaklinks=true,
  colorlinks=true,
  linkcolor=blue,
  anchorcolor=black,
  citecolor=blue,
  urlcolor=black,
}

\newcommand{\firstpagenote}{%
  \AddToShipoutPictureFG*{%
    \AtPageUpperLeft{%
      \hspace*{1.8cm}%
      \raisebox{-1.5cm}{%
        \parbox[t]{0.55\textwidth}{%
          \raggedright\sffamily\small
          Accepted for publication in \href{https://imstat.org/journals-and-publications/statistical-science/}
          {\textcolor{blue}{Statistical Science}}
        }%
      }%
    }%
  }%
}

\begin{document}
\firstpagenote
\maketitle

\begin{abstract}
Replication of scientific studies is important for assessing the credibility of their results. However, there is no consensus on how to quantify the extent to which a replication study replicates an original result. We propose a novel Bayesian approach for replication studies based on mixture priors.
The idea is to use a mixture of the posterior distribution based on the original study and a non-informative distribution as the prior for the analysis of the replication study. The mixture weight then determines the extent to which the original and replication data are pooled.
Two distinct strategies are presented: one with fixed mixture weights, and one that introduces uncertainty by assigning a prior distribution to the mixture weight itself. Furthermore, it is shown how within this framework Bayes factors can be used for formal testing of relevant scientific hypotheses, such as tests on the presence or absence of an effect or whether the mixture weight equals zero (completely discounting the original data) or one (fully pooling with the original data). To showcase the practical application of the methodology, we analyze data from three replication studies. Our findings suggest that mixture priors are a valuable and intuitive alternative to other Bayesian methods for analyzing replication studies, such as hierarchical models and power priors. We provide the free and open source R package \texttt{repmix} that implements the proposed methodology.
  \\[1ex]
\end{abstract}



\section{Introduction} \label{Introduction}

The integrity and credibility of scientific research heavily relies on the replicability of its results \citep{NSF2019}. However, in recent years, an increasing number of published findings failed to replicate, leading to growing concerns about a ``replication crisis'' in several scientific fields.

For instance, the Reproducibility Project: Psychology \citep{open_science_2015} demonstrated that only about $36\%$ of $97$ psychological studies could be replicated in terms of statistical significance. Similarly, large-scale replication projects in economics and social science reported replication rates around $60$ to $70\%$ \citep{Camerer_2016, Camerer_2018}, while attempts in preclinical cancer-biology studies reported replication rates around $40\%$ \citep{Errington_2021}.
As a consequence, there is increasing emphasis within the scientific community on the importance of replication studies \citep{NWO2016, Nature2022}. 


Typically, a distinction is made between direct and conceptual replications \citep{schmidt2009shall, nosek2017making}. Direct replications attempt to replicate a previously observed result using new data collected according to a protocol that closely mirrors the original study -- often increasing the sample size to ensure sufficient power to corroborate the
original finding. Conceptual replications test the generalizability and robustness of a finding by changing aspects of the study design, such as the sample or the experimental conditions. In this paper, we will focus on direct replications, as they are an important first step towards verifying the reliability of a scientific finding before assessing its generalizability \citep{Simons2014}.

Establishing the success of a replication remains a challenging task. Multiple statistical methodologies, ranging from frequentist to Bayesian paradigms and even hybrid models of both, have been suggested to quantify the degree of success a replication study achieved in replicating the original result \citep[among others]{bayarri_2002a,bayarri_2002b, verhagen2014bayesian, Patil_2016, Johnson_2017, Harms_2019, Hedges_2019, Held_2020, mathur_2020, Pawel_Held_2020, pawel_held_2022, held_micheloud_2022, micheloud2023assessing, pawel2023power}. 

A common approach is to compare the original and replication effect sizes, where if the replication yields a substantially smaller effect estimate, the credibility of the original study result decreases. 
 A variation of this is to investigate whether the replication effect estimate is consistent with the original effect estimate using hypothesis tests. Meta-analytic approaches pool original and replication results to produce a combined effect estimate. However, such symmetric approaches implicitly assume exchangeability between studies -- an assumption that may be inappropriate when one is explicitly testing the credibility of an earlier finding, as studies are often conducted under different standards. This arises not only from methodological considerations but also from how researchers cognitively evaluate evidence.  For instance, \citet{ernst2018researchers} found that researchers exhibit an unexpected recency anchoring effect, placing greater weight on the most recently presented evidence -- particularly when replication attempts fail to confirm the initial finding. 
Consequently, asymmetric methods that treat the original and replication studies differently are often more suitable to quantify replication success \citep{Held_2020, koppe2025assessing}. 
\color{black}

\subsection{Bayesian methods for replication studies}
Analyzing replication studies involves per definition the use of historical data -- the data from the original study. Given the inherent nature of sequential information updating, Bayesian methods are natural tools for this purpose. Consequently, an intuitive way to incorporate historical information is to use a prior distribution based on the data from the original study for the analysis of the replication data. In its simplest form, one could use the posterior distribution of the model parameters based on the original data as the prior for the replication analysis. However, this may be problematic if there is heterogeneity between the two studies, as the resulting posterior may then conflict with both studies. This is of particular concern in the replication setting, where replication studies often show less impressive effects than their original counterparts \citep[e.g., the effect estimates in][were on average only half as large as the original ones]{open_science_2015}, often argued to happen because of stricter control of biases and researcher degrees of freedom, for example, via preregistration of the replication study.

Building on the long tradition of Bayesian hierarchical modeling and meta-analysis \citep[among others]{dempster1977maximum, rubin1981estimation, Hedges1985, smith1995bayesian, sutton2000methods, spiegelhalter2004bayesian}, recent methodological developments have proposed more sophisticated approaches to mitigate potential conflict between historical and current data, and ``borrow information'' from the historical data in an adaptive way \citep[for an overview see e.g.,][]{Lesaffre2024}.
Notably, power priors \citep{chen_ibrahim2000, duan2006}, hierarchical models \citep{thall_et_all_2003,Berry_2013}, and mixture priors \citep{Schmidli_2014, Yang_2023} are three prominent approaches in this domain. The power prior, in its basic version, is derived by updating an initial prior distribution with the likelihood of the historical data raised to the power parameter $\alpha$, ranging between zero and one, which determines the degree to which historical data influence the prior distribution. Power priors evaluate two primary concepts of successful replication \citep{pawel2023power}. Firstly, they ensure that the replication study confirms the presence of a tangible effect, often by assessing the effect size $\theta$, and checking if it differs significantly from zero. Secondly, they assess how well the original data matches with the replication data, as an $\alpha$ value close to one means both studies align seamlessly, while a value close to zero implies a disagreement between the original and the replication study.

Hierarchical modeling offers an alternative way to incorporate historical data into Bayesian analyses. The idea is to assume a hierarchical model where the true original $\theta_o$ and replication effect sizes $\theta_r$ are sampled themselves from a distribution around an overall effect size $\theta$. The variance $\tau^2$ of this distribution then determines the similarity between the studies, a value of zero corresponding to identical true effects while a large value corresponds to heterogeneity. Works by \citet{bayarri_2002a,bayarri_2002b} and \citet{Pawel_Held_2020} have effectively applied this approach in replication scenarios.

Mixture priors represent yet another way to 
adaptively borrow information from historical data \citep{OHagan_2012, egidi_et_al_2022, zwet2022proposal}.
Essentially, a mixture prior combines a prior based on the historical data with a non-informative one, allocating distinct mixing weights to each component. The informative prior encourages information borrowing, while the non-informative prior indicates limited or no use of historical information.
The robust meta-analytic predictive (MAP) prior presented by \citet{Schmidli_2014}, which mixes a MAP prior derived from multiple historical studies with a non-informative prior, is an example of a mixture prior used for historical data borrowing. 
In the replication setting, \citet{consonni2023assessing}
have proposed a mixture prior modification of the reverse-Bayes method from \citet{pawel_held_2022} to limit prior-data conflict between the original study and a ``sceptical prior'' that is used to challenge it. The set of 21 replication studies from the Social Sciences Replication Project have also been jointly analyzed with a Bayesian mixture model to estimate an overall true positive rate and an effect size deflation factor \citep[see also \url{https://osf.io/nsxgj}]{Camerer_2018}. However, apart from these two works, mixture prior modeling has not been applied to replication studies in any way, particularly not in its most basic form of using a mixture prior based on the original study for the analysis of the replication study.

The aim of this paper is therefore to present a novel and conceptually intuitive Bayesian approach for quantifying replication success based on mixture priors. The idea is to use a mixture of the posterior distribution based on the original study and a non-informative distribution as the prior for the analysis of the replication study. The mixture weight then determines the extent to which the original and replication data are pooled. This methodology is illustrated using data from three replication studies, which were part of the replication project from \citet{ebersole2016many}, detailed in the following Section \ref{Running example}. Section~\ref{Mixture prior modeling of replication studies} then describes the process of deriving mixture priors from data of an original study within a meta-analytic framework, presenting a general approach for integrating original data into the mixture prior. In this exploration, two distinct approaches are examined: the first fixes the mixture weights, while the second introduces uncertainty by assigning a prior distribution on the mixture weight parameter. Section \ref{posterior checks} presents posterior predictive checks to assess whether the observed data are
consistent with the assumed Bayesian model. In Section \ref{Hypothesis testing} different hypotheses regarding the underlying parameters of interests are examined. Bayes factors are derived offering a quantitative measure of evidence for one hypothesis over another. A comparison of our mixture prior framework with established approaches, particularly hierarchical models and power priors, is illustrated in Section \ref{sec:comparison}.
Finally, Section \ref{Discussion} provides concluding remarks, emphasizing the strengths and limitations of the mixture prior approach, along with insights into potential extensions.

\subsection{Running example} \label{Running example}

We examine an experiment from the "Many Labs 3" replication project \citep{ebersole2016many}. Specifically, twenty university labs across the United States and Canada collected data following a standardized protocol to replicate a classical social psychology experiment on the theory of "moral credentialing" (Study 1 from \cite{monin2001moral}). According to this theory, individuals who have been given an initial opportunity to demonstrate that they are not prejudiced -- and thereby establish "moral credentials" -- are subsequently more likely to exhibit prejudiced attitudes. This may occur because their initial non-prejudiced behavior serves as justification for subsequent behavior that might otherwise appear prejudiced.

\begin{figure*}[htb!]
    \centering
        \includegraphics[width=\textwidth, height=8cm]{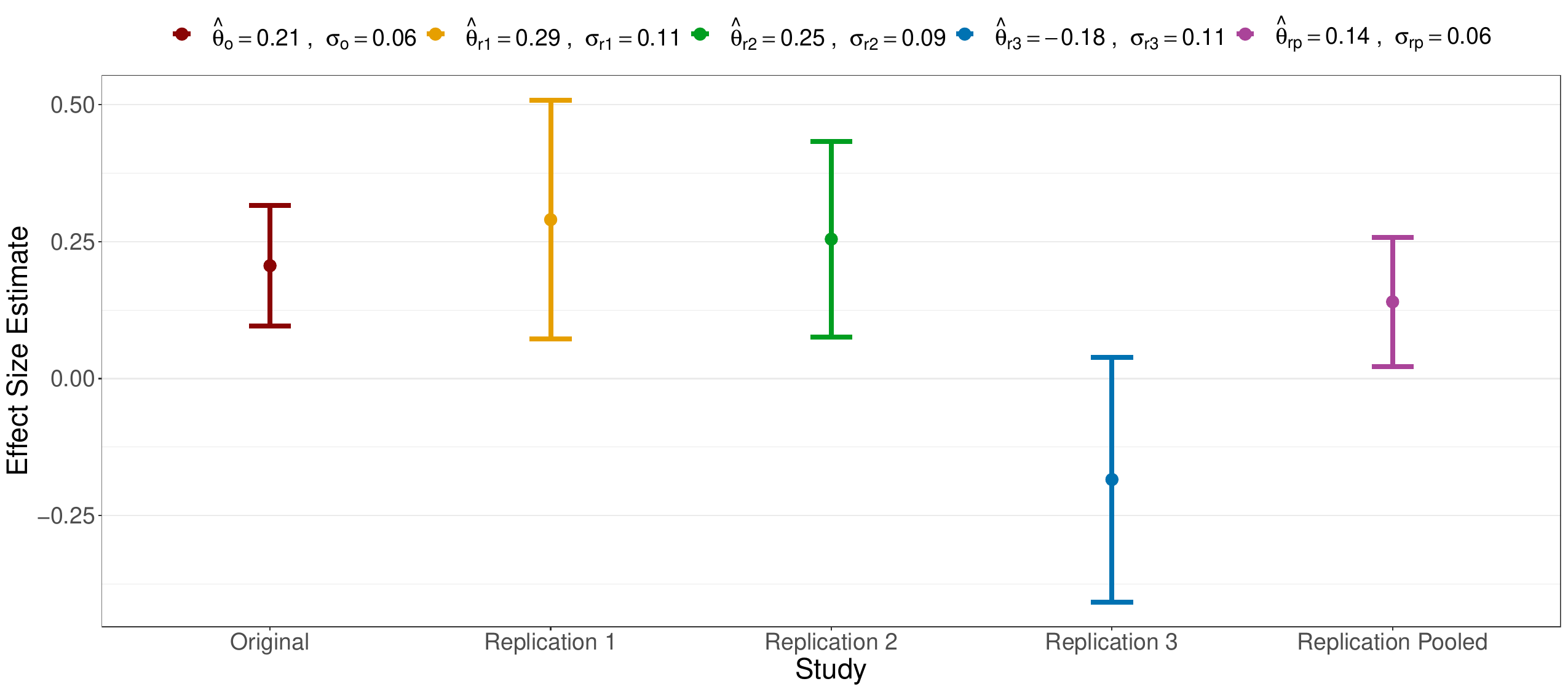}
        \caption{Effect size estimates (on Fisher's z-scale) and $95\%$ CI for the ``Moral credentialing'' original study, the three independent replications, and the pooled replication.}
        \label{Effect_Size_Plot}
\end{figure*}

The original study found evidence for this hypothesis. Its main finding was drawn from a sample of $202$ participants, which led to an effect size estimate of $\hat{\theta}_o = 0.21$ with standard error $\sigma_o = 0.06$. As part of the Many Labs 3 project, the "moral credentialing" experiments were replicated in twenty independent university laboratories using identical procedures, with results revealing both a smaller average effect and showing signs of inter-site variation beyond sampling error \citep{ebersole2016many}. To investigate these findings more concretely and assess the robustness of our proposed modeling approach, we focus on three direct replications that exemplify different degrees of inter-site variability. The first replication, conducted at the University of Toronto, yielded a relatively large effect estimate of $\hat{\theta}_{r_1} = 0.29$ with $\sigma_{r_1} = 0.11$. The second, carried out at Montana State University, produced a comparable estimate, $\hat{\theta}_{r_2} = 0.25$ with $\sigma_{r_2} = 0.09$. In contrast, the third replication, conducted at Ashland University, reported a substantially smaller -- and directionally opposite -- effect estimate of $\hat{\theta}_{r_3} = -0.18$ with $\sigma_{r_3} = 0.11$. Figure \ref{Effect_Size_Plot} shows the effect size estimates along with their $95\%$ confidence intervals for the original study, its three independent replications, and the pooled replication. All reported effect size estimates are presented on Fisher's $z$-scale to ensure approximate normality, as recommended by \citet{mathur_2020}.

\section{Mixture prior modeling of replication studies} \label{Mixture prior modeling of replication studies}
In the following, we use a meta-analytic framework which can be applied to a broad range of data types and effect sizes \citep[see e.g., chapter 2.4 in][]{spiegelhalter2004bayesian}. Define by $\theta$ the unknown effect size, with $\hat{\theta}_o$ and $\hat{\theta}_{r_i}$ being the estimated effect size from the original study and replication $i = 1, \dots, m$, respectively. As in meta-analyses \citep{Hedges1985}, it is common in replication study settings to specify that the likelihood of the effect size estimates is approximately normal \citep[among others]{Hedges2019b,mathur_2020,Held_2020, pawel_held_2022, pawel2023power,micheloud2024replication, Xiao2024}, which is reasonable when replication studies have sufficiently large sample sizes \citep{sutton2000methods}. That is 
\begin{align*}
    &\hat{\theta}_o \mid \theta \sim \mathrm{N}\left(\theta, \sigma_o^2\right)& 
    &\hat{\theta}_{r_i} \mid \theta \sim \mathrm{N}\left(\theta, \sigma_{r_i}^2\right),
\end{align*}
where $\sigma_\cdot$ represents the standard error of an estimate, which is assumed to be known. There are circumstances under which the effect size might need a particular transformation, such as a logit function or a log function transformation, to refine the normal distribution approximation. Additionally, adjusting the effect size for confounders via regression might also be necessary.
Finally, define the pooled replication effect size estimate and its standard error by
\begin{align*}
    &\hat{\theta}_{r_p} = 
    \frac{\sum_{i=1}^m \hat{\theta}_{r_i}/\sigma^2_{r_i}}{\sum_{i=1}^m 1/\sigma^2_{r_i}}&
    \sigma_{r_p} = \sqrt{\frac{1}{\sum_{i=1}^m 1/\sigma^2_{r_i}}},
\end{align*}
which are sufficient statistics for inference regarding the effect size parameter $\theta$, that is, we have that the likelihood of a sample of independent replication studies is
\begin{align*}
    \prod_{i=1}^m \mathrm{N}(\hat{\theta}_{r_i} \mid \theta, \sigma^2_{r_i})
    = K \times \mathrm{N}(\hat{\theta}_{r_p} \mid \theta, \sigma^2_{r_p}),
\end{align*}
with $\mathrm{N}(\cdot \mid m,v)$ the normal density function with mean $m$ and variance $v$ and $K$ a constant that does not depend on the effect size $\theta$.
In the following, we will investigate posterior distribution and Bayes factor analyses related to the effect size $\theta$ and based on the likelihood of the pooled replication effect size estimate and standard error $\hat{\theta}_{r_p} \mid \theta \sim \mathrm{N}(\theta, \sigma^2_{r_p})$. For both analyses, the constant $K$ cancels out and the approach thus encompasses both the analysis of a single replication study ($m = 1$ so that $\hat{\theta}_{r_p} = \hat{\theta}_{r_1}$ and $\sigma_{r_p} = \sigma_{r_1}$) or multiple replication studies ($m > 1$).

The aim is now to develop a mixture prior for the effect size $\theta$ that combines two distinct components. The first component is derived from the original study, akin to the meta-analytic-predictive (MAP) prior described by \citet{spiegelhalter2004bayesian} and \citet{Neuenschwander_2010}; and the second component is a normal prior that provides a robust alternative in case there is conflict between the replication and original data \citep{Schmidli_2014}. In detail, it is 
\begin{equation} \label{eqn:1}
    \pi(\theta \mid \hat{\theta}_o, \omega) = \omega \mathrm{N}(\theta \mid \hat{\theta}_o,\sigma^2_o) + (1-\omega) \mathrm{N}(\theta \mid \mu,\tau^2).
\end{equation}
The mean $\mu$ and variance $\tau^2$ of the alternative are typically specified such that the prior is proper but non-informative (e.g., $\mu = 0$ and $\tau^2$ large).
Clearly, by setting $\omega = 1$, we obtain a prior that leads to a complete pooling of the data from both studies, while setting $\omega = 0$ completely discounts the original data. For a $0 < \omega < 1$, there is a gradual compromise between these two extremes.

In a mixture prior as in (\ref{eqn:1}), setting an appropriate mixing weight $\omega$, is a complex but crucial task. It is essential that the chosen $\omega$ accurately reflects the level of agreement between the original and replication studies. A prior that places too much weight towards the non-informative component can undermine the effectiveness of borrowing from the original study, leading to an underestimate of the real agreement between the original and replication studies. On the contrary, a mixing weight skewed heavily towards the informative prior may result in overestimating the confidence on the similarity between the two studies, introducing a potential bias. 
In the following, we will discuss two strategies for determining the value of $\omega$. The first strategy involves fixing $\omega$ on a predetermined value that is considered reasonable, while the second employs an additional prior specification by taking $\omega$ as a random quantity.

\subsection{Fixed weight parameter} \label{Fixed weight parameter}
After observing the replication data, the mixture prior (\ref{eqn:1}) is updated yielding the posterior distribution 
\begin{equation} \label{eqn:2}
\begin{aligned}
\pi(\theta \mid \hat{\theta}_o, \hat{\theta}_r, \omega)  
&= \dfrac{\mathrm{N}(\hat{\theta}_r \mid \theta, \sigma^2_r) \left\{\omega \mathrm{N}(\theta \mid \hat{\theta}_o,\sigma^2_o) + (1-\omega) \mathrm{N}(\theta \mid \mu,\tau^2)\right\}}{f(\hat{\theta}_r \mid \hat{\theta}_o, \omega)},
\end{aligned}
\end{equation}
the marginal likelihood in the denominator is
\begin{equation} \label{eqn:3}
    \begin{aligned}
    f(\hat{\theta}_r \mid \hat{\theta}_o, \omega) 
    = \omega \mathrm{N}(\hat{\theta}_r\mid \hat{\theta}_o,\sigma_r^2+\sigma_o^2) + (1-\omega)\mathrm{N}(\hat{\theta}_r \mid \mu, \sigma^2_r + \tau^2).
    \end{aligned}
\end{equation}
The closed-form solution \eqref{eqn:3} for the marginal likelihood can be used to show that the posterior is again a mixture of normals
\begin{align*}
     \pi(\theta \mid \hat{\theta}_o, \hat{\theta}_r, \omega) = \omega^\prime \mathrm{N}(\theta \mid m_1, v_1) + (1 - \omega^\prime) \mathrm{N}(\theta \mid m_2, v_2),
\end{align*}
with updated means and variances
\begin{align*}
    &m_1 = (\hat{\theta}_o/\sigma^2_o + \hat{\theta}_r/\sigma^2_r) \times v_1,& &v_1 = (1/\sigma^2_o + 1/\sigma^2_r)^{-1},& \\
    &m_2 = (\mu/\tau^2 + \hat{\theta}_r/\sigma^2_r) \times v_2,& &v_2 = (1/\tau^2 + 1/\sigma^2_r)^{-1},& 
\end{align*}
and updated weight
\begin{align*}
    &\omega^\prime = \left\{1 + \frac{1 - \omega}{\omega} \times \frac{\mathrm{N}(\hat{\theta}_r \mid \mu, \tau^2 + \sigma^2_r)}{\mathrm{N}(\hat{\theta}_r \mid \hat{\theta}_o, \sigma^2_o + \sigma^2_r)}\right\}^{-1}.
\end{align*}
The two posterior components thus represent two ordinarily updated normal posteriors, while the initial weight along with the relative predictive accuracy of the replication data under either component determines the updated weight. The fact that for a fixed mixture weight, the posterior distribution is again a mixture distribution is known from general Bayesian theory \citep{bernardo1994bayesian, spiegelhalter2004bayesian,fruhwirth2019handbook,neuenschwander2023fixed}. The mixture representation of the posterior also shows that the non-informative component has to be proper ($\tau^2 < \infty$) to enable borrowing, as otherwise the updated weight will be $\omega^\prime = 1$, leading always to a complete pooling with the historical data regardless of conflict.

There are different approaches for specifying the mixture weight $\omega$. A straightforward approach involves assigning to $\omega$ a value that is reasonable, based on domain-expert knowledge, regarding the agreement between the two studies. Alternatively, the empirical Bayes estimate of $\omega$ may be used, which represents the value that maximizes the marginal likelihood function~\eqref{eqn:3}. Finally, in order to assess prior sensitivity, a reverse-Bayes approach \citep{good_1950, Best2021, held_reverse_bayes} may be used to find the mixture weight such that a certain posterior distribution is obtained.

\begin{figure*}[htb!]
    \centering
        \includegraphics[width=\textwidth, height=8cm]{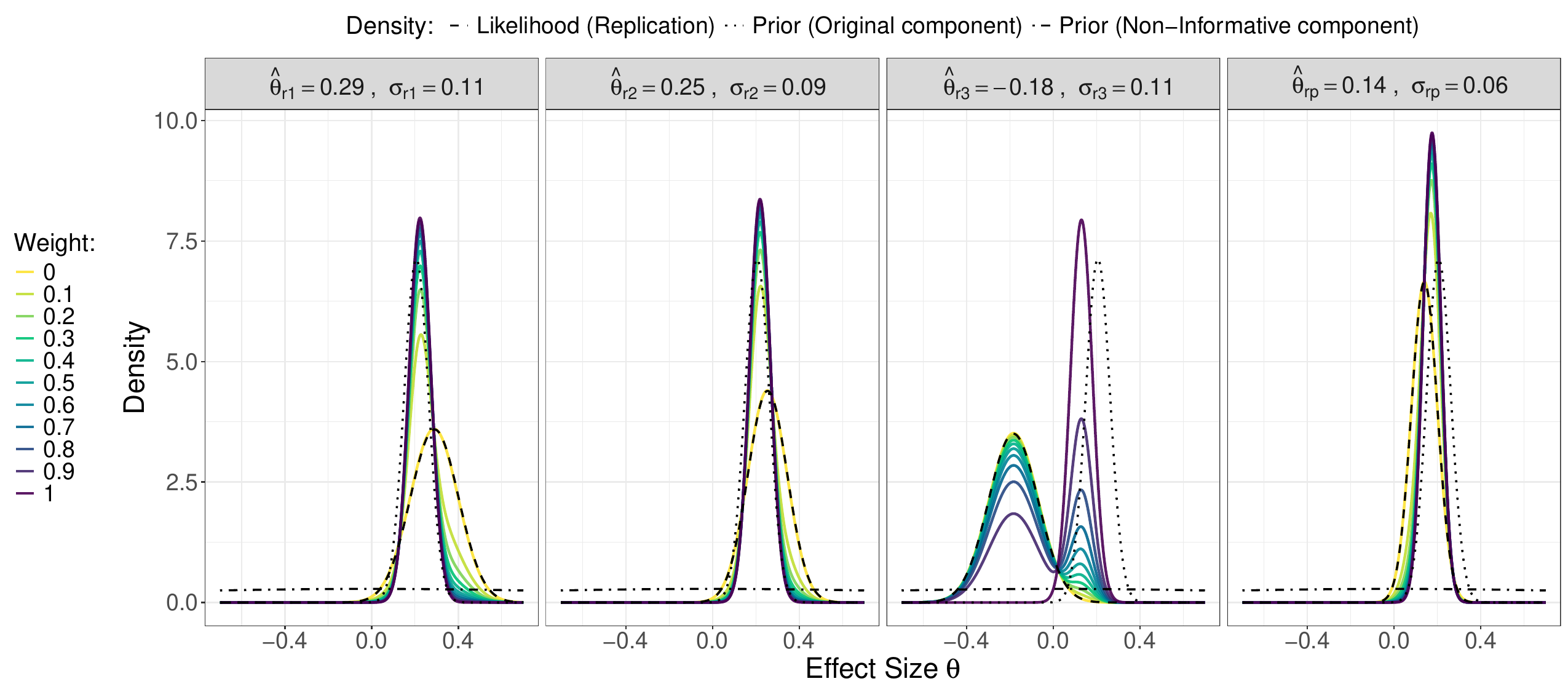}
        \caption{``Moral credentialing'' experiment. Each colored line represents the posterior distribution of effect size $\theta$ under different fixed weights. The black dotted line indicates the prior distribution based on the original study. The dash-dotted line represents a non-informative prior. The dashed line corresponds to the likelihood based on replication data.}
        \label{Posterior_fixed_weights}
\end{figure*}

Returning to the ``Moral credentialing'' experiment from \citet{ebersole2016many} introduced in Section \ref{Running example}, Figure \ref{Posterior_fixed_weights} acts as a sensitivity check showing the shifts in the posterior distribution for the effect size (\ref{eqn:2}) under different fixed weights assigned to the mixture prior. Here, the non-informative prior component in (\ref{eqn:1}) is constructed to be a unit-information normal distribution centred at a mean $\mu = 0$ and with variance $\tau^2 = 2$. A unit-information prior \citep{kass_wasserman_1995}  is chosen to provide only a minimal amount of information. Essentially, its variance is set so that the prior's information has a content equivalent to a unit sample size. The use of unit-information prior is illustrated in several studies such as \citet{NTZOUFRAS2003} for binary response models, \citet{OVERSTALL2010} apply this principle to generalized linear mixed models, and \citet{bove_held} demonstrate its application in the context of generalized linear models. For further details, see also \citet{consonni_etal_2018}. We see that, varying $\omega$, within the range from 0 to 1, induces a progressive transformation of the posterior distribution. At $\omega = 0$, the posterior distribution virtually aligns with the likelihood of the replication study as the influence of the non-informative component is minimal compared to the replication data. Conversely, as $\omega$ increases towards $1$, the posterior distribution gradually becomes more influenced by the prior associated with the original study, leading to a posterior that lies somewhere in between the replication and original likelihood, as the replication borrows information from the original study. 

\begin{figure*}[htb!]
    \centering
        \includegraphics[width=\textwidth, height=8cm]{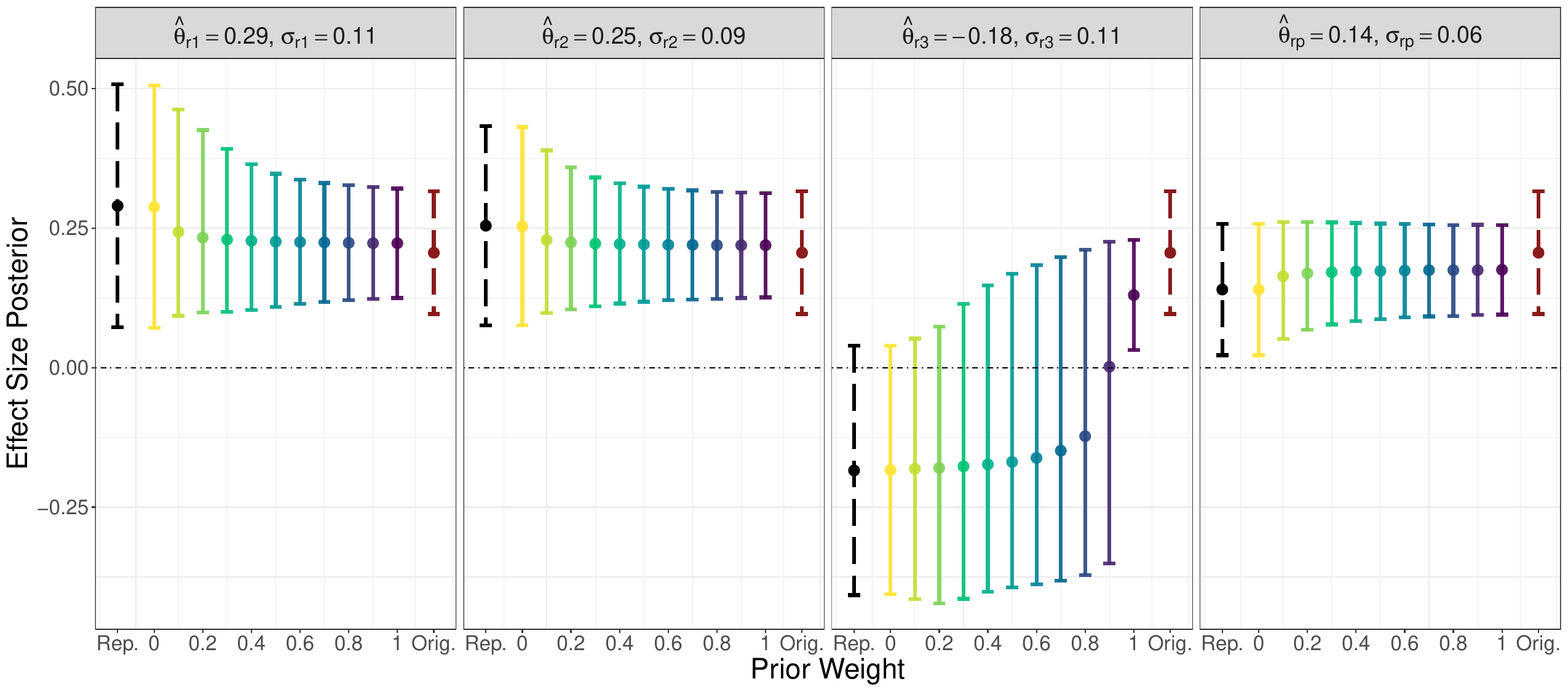}
        \caption{``Moral credentialing'' experiment. Posterior median (points) and $95\%$ highest posterior density interval (HPDI) of the effect size posterior against mixture prior weight assigned to the original study component. On the left and right side of each panel, the corresponding replication study effect estimate and the original study effect estimate with $95\%$ confidence interval.}
        \label{hpdi_fixed_weights}
\end{figure*}

Based on the reverse-Bayes approach \citep{Best2021, held_reverse_bayes}, a tipping-point analysis is conducted to assess the influence of the mixing weight $\omega$ on the resulting posterior distribution. This analysis focuses on the question: ``How much does the mixing weight have to change for the conclusion of the analysis to change?'' Figure \ref{hpdi_fixed_weights} shows the posterior median and the $95\%$ highest posterior density interval (HPDI) of the effect size for different prior weights $\omega \in \{0, 0.1, 0.2, ..., 1\}$ associated with the original study component in (\ref{eqn:1}).  We see that the first, second, and pooled replication scenarios are robust with respect to the choice of weights, as the effect size posterior median and its corresponding $95\%$ HPDI remain substantially above zero across all prior weights, thereby suggesting robust evidence for a genuine effect. In contrast, the third replication is less stable, as the $95\%$ HPDI includes zero up to about a weight of at least 0.9. 
Specifically, this third replication can only be considered as providing evidence for a genuine effect if a mixture weight of at least $0.996$ is chosen, as the replication study alone (i.e., a mixture weight of zero) fails to do so. 

It is important to note that the posterior median can be a misleading point estimate in the case of bimodality. 
In our analysis, only the posterior of the third replication shows a clearly bimodal posterior distribution for certain weight values, while the posteriors of the remaining replications appear unimodal. However, if assessing bimodality by looking at the posterior density is not possible, it may be advisable to at least compute numerical summaries that quantify the potential bimodality \citep[see e.g. sections 2.2-2.10 in][]{OHagan2004}.

\subsection{Prior on the weight parameter} \label{Prior on the weight parameter}
We now introduce a fully specified Bayesian approach for the mixture prior in (\ref{eqn:1}) assuming uncertainty on the weight $\omega$. This approach considers $\omega$ as a random quantity, requiring the specification of a prior distribution $\pi(\omega)$. A natural choice is a Beta distribution 
\begin{equation*}
\omega \mid \eta, \nu \sim \mathrm{Beta}(\eta,\nu), 
\end{equation*}
since $\omega$ is a proportion.
Consequently, this formulation leads to the joint prior distribution for the effect size $\theta$ and the weight $\omega$
\begin{equation} \label{eqn:4}
\begin{aligned}
    \pi(\theta,\omega \mid \hat{\theta}_o, \eta, \nu) 
    &= \pi(\omega \mid \eta, \nu ) \pi(\theta \mid \omega, \hat{\theta}_o) \\
    &= \mathrm{Beta}(\omega \mid \eta, \nu) \left\{\omega \mathrm{N}(\theta \mid \hat{\theta}_o, \sigma^2_o)+ (1-\omega)\mathrm{N}(\theta \mid \mu, \tau^2)\right\},
\end{aligned}
\end{equation}
where $\mathrm{Beta}(\cdot \mid \eta, \nu)$ is the Beta density function with the strictly positive shape parameters $\eta, \nu > 0$. 
Given the joint prior distribution (\ref{eqn:4}) and in light of the replication data, the joint posterior distribution is then
\begin{equation} \label{eqn:5}
\begin{aligned}
    \pi(\theta,\omega \mid \hat{\theta}_r,\hat{\theta}_o, \eta, \nu)  
    &= \dfrac{\mathrm{N}(\hat{\theta}_r \mid \theta, \sigma^2_r) \mathrm{Beta}(\omega \mid \eta, \nu) \left\{\omega \mathrm{N}(\theta \mid \hat{\theta}_o, \sigma^2_o)+ (1-\omega)\mathrm{N}(\theta \mid \mu, \tau^2)\right\}}{f(\hat{\theta}_r \mid \hat{\theta}_o, \eta,\nu)}.
\end{aligned}
\end{equation}
The marginal likelihood in the normal mixture model with random weights can be determined through a closed-form solution, similar to Equation (\ref{eqn:3}). In this scenario, it depends on the expected value of the weight parameter $\omega$ and on the updated normal prior components of the mixture 
\begin{equation*}
    \begin{aligned}
        f(\hat{\theta}_r \mid \hat{\theta}_o, \eta,\nu) 
        &= 
        \int\int \mathrm{N}(\hat{\theta}_r \mid \theta, \sigma^2_r) \mathrm{Beta}(\omega \mid \eta, \nu) \left\{\omega \mathrm{N}(\theta \mid \hat{\theta}_o, \sigma^2_o)+ (1-\omega)\mathrm{N}(\theta \mid \mu, \tau^2)\right\} \mathrm{d} \theta \mathrm{d} \omega \\
        &= \int\mathrm{Beta}(\omega \mid \eta, \nu)  \left\{ \omega \mathrm{N}(\hat{\theta}_r\mid \hat{\theta}_o,  \sigma_r^2+\sigma_o^2) + (1-\omega)\mathrm{N}(\hat{\theta}_r \mid \mu, \sigma^2_r + \tau^2)\right\} \mathrm{d}\omega \\
        &= \left(\dfrac{\eta}{\eta+\nu}\right)  \left\{\mathrm{N}(\hat{\theta}_r \mid \hat{\theta}_o, \sigma^2_r +\sigma^2_o)-\mathrm{N}(\hat{\theta}_r \mid \mu, \sigma^2_r + \tau^2)\right\} + \mathrm{N}(\hat{\theta}_r \mid \mu, \sigma^2_r+ \tau^2).
    \end{aligned}
\end{equation*}
Consequently, in the case of a random weight, the marginal likelihood is similar to that in Equation (\ref{eqn:3}), with the difference of replacing the fixed weight with the expected weight over the prior. 
By integrating $\theta$ out in (\ref{eqn:5}), the marginal posterior distribution of $\omega$ can be expressed as
\begin{equation} \label{eqn:6}
\begin{aligned}
    \pi(\omega \mid \hat{\theta}_r,\hat{\theta}_o, \eta, \nu) 
    &=\dfrac{\mathrm{Beta}(\omega \mid \eta, \nu)  \left\{ \omega \mathrm{N}(\hat{\theta}_r \mid \hat{\theta}_o, \sigma^2_r +\sigma^2_o) + (1-\omega)\mathrm{N}(\hat{\theta}_r \mid \mu, \sigma^2_r + \tau^2)\right\}}{f(\hat{\theta}_r \mid \hat{\theta}_o, \eta,\nu)}. 
\end{aligned}
\end{equation}
The marginal posterior of $\theta$ is given by
\begin{equation*}
\begin{aligned}
   \pi(\theta \mid \hat{\theta}_r,\hat{\theta}_o, \eta, \nu) 
   &=  \dfrac{\mathrm{N}(\hat{\theta}_r \mid \theta, \sigma^2_r)  \left(\dfrac{\eta}{\eta+\nu}\right)\left\{\mathrm{N}(\theta \mid \hat{\theta}_o,\sigma^2_o)-\mathrm{N}(\theta \mid \mu, \tau^2)\right\}+\mathrm{N}(\theta \mid \mu, \tau^2)}{f(\hat{\theta}_r \mid \hat{\theta}_o, \eta,\nu)}. 
\end{aligned}
\end{equation*}
In summary, when introducing uncertainty in the mixture weight $\omega$ via a Beta prior, the marginal likelihood of the data, the joint and marginal posteriors of the effect size $\theta$, and the mixture weight $\omega$ are still available in closed-form. Moreover, the marginal likelihood and marginal posterior of $\theta$ are of the same form as with a fixed mixture weight $\omega$ as shown in the previous section, but with $\omega$ replaced by its expected value under its prior distribution \citep{neuenschwander2023fixed}.

\begin{figure*}[htb!]
    \centering
        \includegraphics[width=\textwidth, height=8cm]{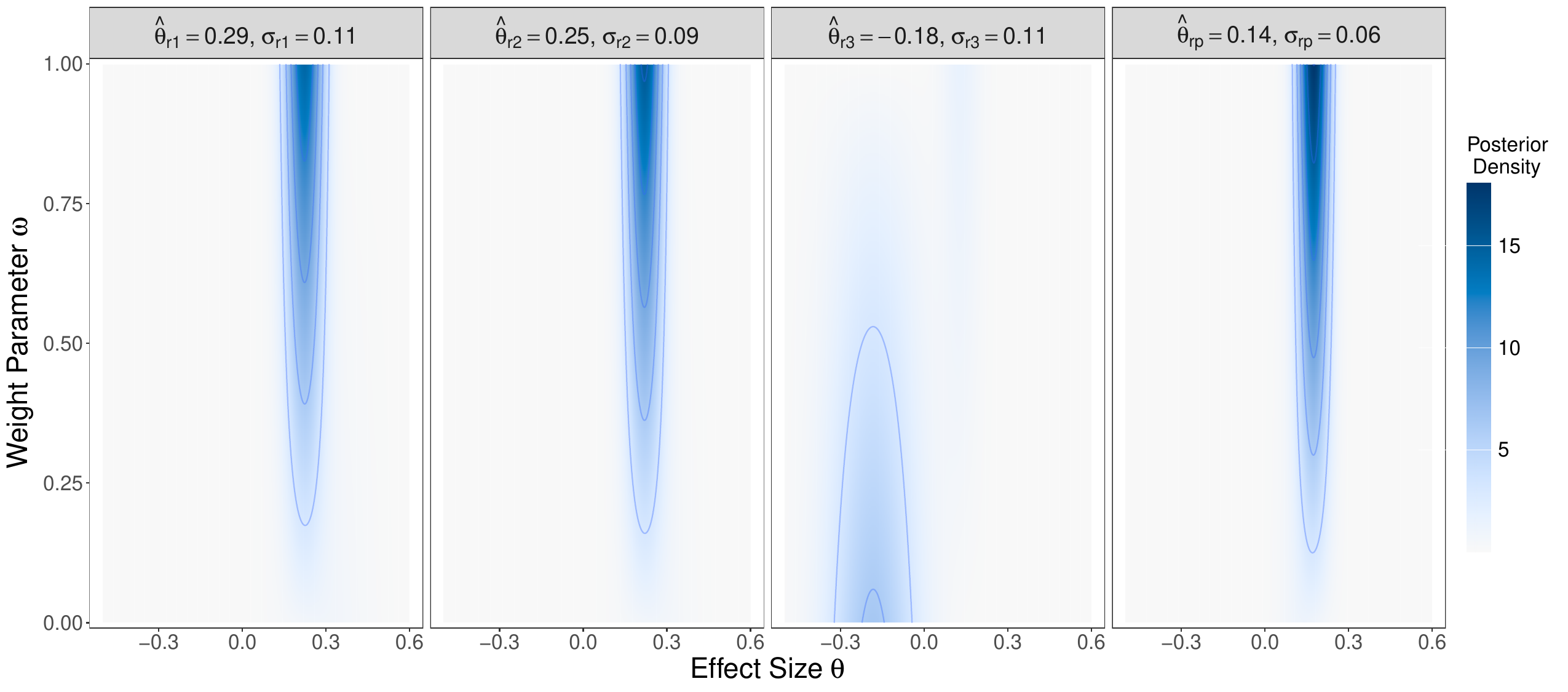}
        \caption{Joint posterior distribution of the effect size $\theta$ and the weight parameter $\omega$ considering the data from the ``Moral credentialing'' experiment, its three replications, and the pooled replication.}
        \label{plot_joint}
\end{figure*}

Figure \ref{plot_joint} shows the contour plot of the joint posterior distribution for the effect size $\theta$ and the weight parameter $\omega$ considering the data from the ``Moral credentialing'' experiment, its three replications, and the pooled replication. In our analysis, we employ a mixture prior, as in (\ref{eqn:4}), in which the informative prior component is derived from the original study, while the non-informative prior is a unit-information prior as in Section \ref{Fixed weight parameter}.
Additionally, we adopt a flat prior distribution for the weight parameter choosing a $\mathrm{Beta}(1,1)$.
We see that for the first, second, and pooled replications, the posterior distribution is concentrated around weight parameter values close to one, reflecting the similarity between the original and replication results. In contrast, for the third replication the posterior distribution is concentrated around zero, indicating a conflict between the original study and the results of this replication. In addition, because it is based on three replications instead of just one, the posterior based on the pooled replications is much more peaked than the others.

\begin{figure*}[h!]
    \centering
        \includegraphics[width=\textwidth, height=8cm]{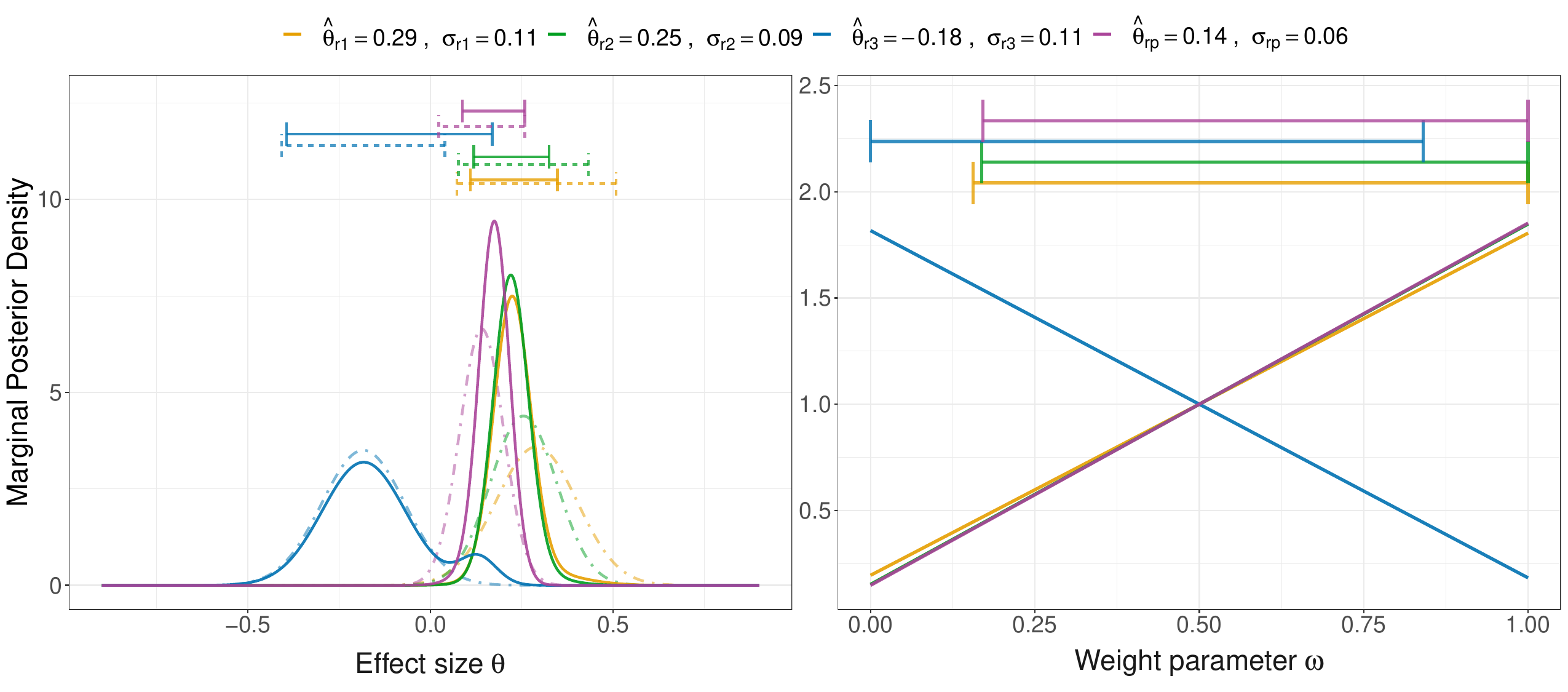}
        \caption{Marginal posterior distributions of the effect size $\theta$ (left) and the weight parameter $\omega$ (right) considering the data from the ``Moral credentialing'' experiment, the chosen three replications, and the pooled replication. The dashed lines represent the posterior density of the effect size $\theta$, derived exclusively from the replication data, without considering the original data, and assuming a uniform prior for the effect size $\pi(\theta)\propto 1$. The horizontal error bars indicate the $95\%$ highest posterior density credible intervals (HPDI).}
        \label{plot_marg_post_joint}
\end{figure*}

Figure \ref{plot_marg_post_joint} shows the marginal posterior distributions for the effect size $\theta$ (left) and the weight parameter $\omega$ (right). The plot related to $\theta$ is enriched by contrasting it with the posterior distribution of $\theta$ based solely on the replication data, represented as a dashed line. This effectively illustrates the added value of integrating the original data through a mixture prior. The blue marginal posterior, corresponding to the most divergent estimate $\hat{\theta}_{r_3} = -0.18$, shows a tendency to incorporate less information among the three replications, leading to a more heavy-tailed posterior distribution with a small "hump" on the right. The discrepancy with the original study increases the variance of the posterior distribution, as is evident when compared with the replication-only posterior shown by the dashed blue line. This is further highlighted in the $95\%$ HPDI, which ranges from $-0.40$ to $0.17$, exceeding the $95\%$ HPDI range of $-0.41$ to $0.04$ observed when the replication data is analyzed without considering the original study, represented by the dashed horizontal blue bar. Conversely, the yellow and green marginal posteriors -- both associated with the most coherent replications $\hat{\theta}_{r_1} = 0.29$ and $\hat{\theta}_{r_2} = 0.25$, respectively -- result in a noticeably narrower $95\%$ HPDI compared to the one derived solely from the replication data. Additionally, the magenta marginal posterior based on the pooled replication $\hat{\theta}_{r_p} = 0.14$ results to be the most peaked density. It is worth noting that these marginal posteriors are equivalent to those obtained when the weight parameter is fixed at $\omega = 0.5$, as shown in Figure~\ref{Posterior_fixed_weights}, because the expected value of a $\mathrm{Beta}(1,1)$ distribution is $0.5$.

The right panel in Figure \ref{plot_marg_post_joint}  shows the marginal posterior distribution for the weight parameter $\omega$, under the assumption of a flat prior
distribution for $\omega$. 
Following the formula as detailed in (\ref{eqn:6}) and under a flat prior, 
this yields a linearly increasing/decreasing posterior density. However, for non-flat priors (i.e., $\mathrm{Beta}(\eta, \nu)$ with $\eta \neq 1$ and $\nu \neq 1$), the posterior density of the weight $\omega$ is not linear anymore. Instead, it is a product of the prior Beta density (which introduces non-linearity whenever the parameters differ from $1$) and a weighted combination of normal densities.
 The first, second, and pooled replications, highlighted in yellow, green, and magenta respectively, display linear marginal posterior distributions that increase monotonically, indicating a peak at $\omega=1$. This suggests compatibility between the two replications and the pooled replication with respect to the original study. Conversely, the linear marginal distribution of the third replication, illustrated in blue, exhibits a monotonically decreasing trend with the most probable value at $\omega = 0$. This trend suggests a notable disagreement between this replication and the original study. Nevertheless, it is worth noting that the HPDIs remain considerably wide across all the replication scenarios,
despite the precisely estimated effect sizes. 

\section{Posterior predictive checking}\label{posterior checks}


Posterior predictive checks (PPCs) \citep{rubin1984bayesianly, gelman1996posterior, gelman2013bayesian} are important tools to evaluate whether the observed data are consistent with the assumed Bayesian model. The basic idea is to simulate new datasets based on the fitted model and then compare the statistics computed from the simulated data with the same statistics computed from the observed data. If these differ substantially, the model is likely misspecified and should be refined.
Specifically, new effect estimates $\hat{\theta}_{r,\text{new}}$ are generated from the posterior predictive distribution
\begin{align*}
    f(\hat{\theta}_{r,\text{new}} \mid \hat{\theta}_r) = \int \pi(\theta \mid \hat{\theta}_{o}, \hat{\theta}_r, \omega) f(\hat{\theta}_{r,\text{new}} \mid \theta)\, d\theta,
\end{align*}
where $\pi(\theta \mid \hat{\theta}_{o}, \hat{\theta}_r, \omega)$ is the posterior distribution for
the effect size and $f(\hat{\theta}_{r,\text{new}} \mid \theta)$ is the likelihood of the hypothetical estimates.

If the model fits well, summary statistics (e.g., the sample mean or standard deviation) from the newly generated estimates should be similar to those from the observed estimate. This can be assessed using posterior predictive $p$-values (PPPs) \citep{guttman1967use,rubin1984bayesianly, meng1994posterior, gelman1996posterior}, defined as the probability that the test statistic in a simulated dataset exceeds that in the observed data
\begin{align*}
    \text{PPP}(\hat{\theta}_r, T) 
= \Pr\left\{ T(\hat{\theta}_{r,\text{new}}) \,\geq\, T(\hat{\theta}_r) \;\middle|\; \hat{\theta}_r \right\}, 
\end{align*}
where $T(\cdot)$ is the test statistic.
It is important to note that PPPs are not classically calibrated. In general, they do not follow a uniform distribution even when the model is correctly specified \citep{bayarri2000p}. Thus, PPPs are usually interpreted as model-diagnostic summaries rather than formal test statistics \citep{gelman2013bayesian}. Values very close to $0$ or $1$ indicate that the observed test statistic is in the tails of the predictive distribution, suggesting that the model does not fit the data well.

\begin{figure*}[h!]
    \centering
        \includegraphics[width=\textwidth, height=8cm]{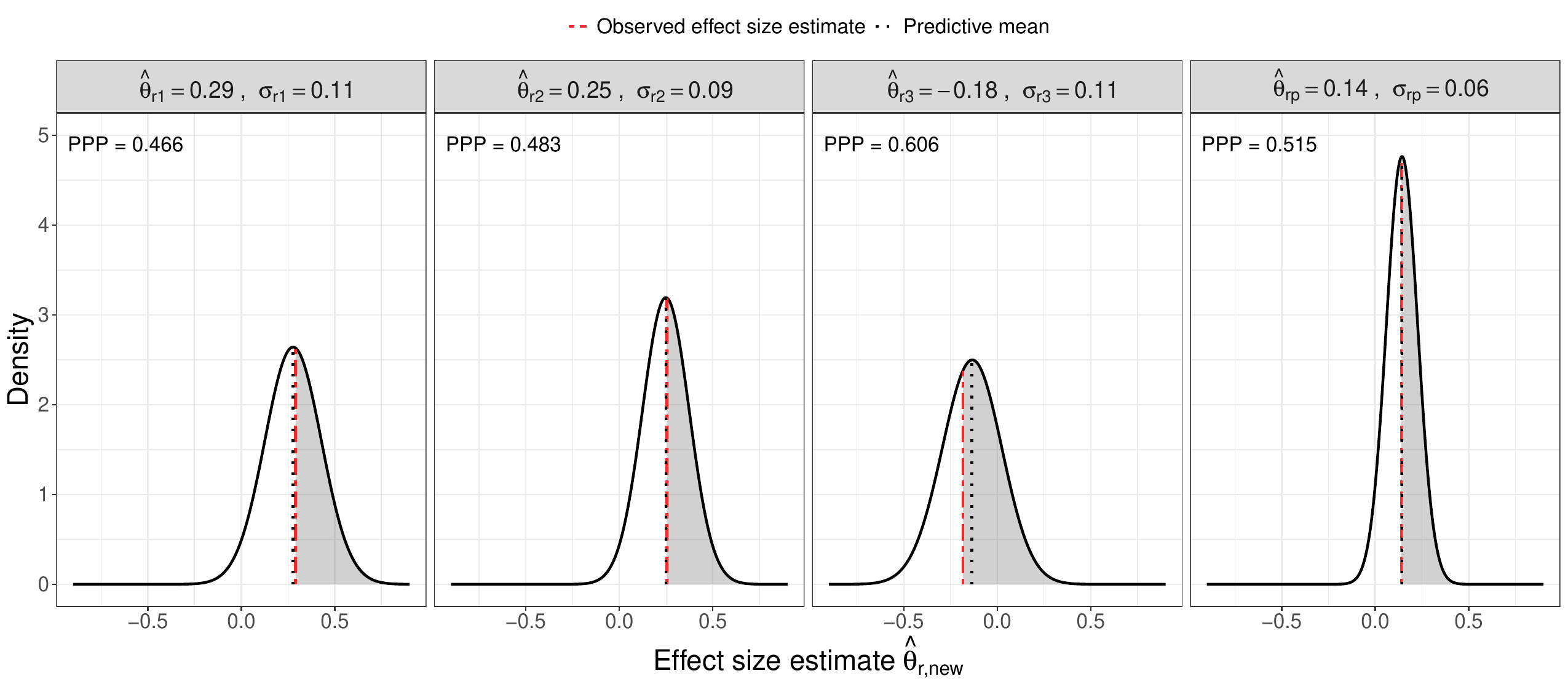}
        \caption{“Moral credentialing” experiment. Posterior predictive checks comparing the observed replication effect size estimate (red dashed line), with the posterior predictive distribution (PPD) of new effect size estimates (black curve). The mean of the PPD is shown as a black dotted line. The shaded grey region denotes values of the distribution more extreme than the observed replication effect size estimate. On the top left corner the corresponding posterior predictive p-value (PPP) for each replication.}
        \label{plot_ppd}
\end{figure*}

In our context, no simulation is needed to compute PPPs as the posterior predictive distribution is analytically tractable and is again a mixture of normals. 
Figure~\ref{plot_ppd} displays the posterior predictive distribution of each replication effect estimate, derived from the posterior distribution shown in Figure~\ref{plot_joint}. In each case, the observed estimate (red dashed line) lies well within the posterior predictive distribution and close to the predictive mean (black dotted line), indicating that there are no evident discrepancies between the model and the observed estimates. Furthermore, it presents PPPs using as test statistic the observed replication effect size estimate $T(\hat{\theta}_r) = \hat{\theta}_r $. They range from $0.466$ to $0.606$ in the three individual replications and in the pooled replication, providing no evidence of misspecification of the Bayesian mixture model.

\section{Hypothesis testing} \label{Hypothesis testing}
Estimating the parameters of a model is one aspect, but in statistical analysis, one may also want to test the plausibility of different scientific hypotheses.  Within the Bayesian framework, the Bayes factor is a key tool for assessing and comparing hypotheses about the parameters \citep[among others]{good_1958, jeffreys1961theory, kass_raftery_1995, GRONAU201780, schonbrodt_bayes_2018, schad2023workflow},
 and Bayes factors have been used for the analysis of replication studies before \citep{Bayarri1991, bayarri_2002a, verhagen2014bayesian, ly2019replication, Harms_2019, pawel_held_2022, VanLissa2024, pawel2023power}.   Although mixture priors have been employed most frequently to compute posterior tail probabilities, their potential use for formal hypothesis testing through Bayes factors has not yet been fully investigated.

Let us consider the replication data (summarized by the effect size estimate $\hat{\theta}_r$) and let $\mathcal{H}_0$ and $\mathcal{H}_1$ be two competing hypotheses. The Bayes factor is then given by the updating factor of the prior odds of the hypotheses to their posterior odds
\begin{align*}
    \mathrm{BF}_{01} 
    = \dfrac{\Pr(\mathcal{H}_0 \mid \hat{\theta}_r)}{\Pr(\mathcal{H}_1 \mid \, \hat{\theta}_r)} \bigg / \dfrac{\Pr(\mathcal{H}_0)}{\Pr(\mathcal{H}_1)}
    = \dfrac{f(\hat{\theta}_r\mid \mathcal{H}_0)}{f(\hat{\theta}_r\mid \mathcal{H}_1)},
\end{align*}
which simplifies to the ratio of marginal likelihoods (or evidences) as shown by the second equality.
As such, the Bayes factor is a quantitative tool to measure the relative evidence that we have for $\mathcal{H}_0$ over $\mathcal{H}_1$. For example, when the a priori probabilities of both hypotheses are assumed to be equal, a Bayes factor greater than one indicates that the data are more likely under $\mathcal{H}_0$ than $\mathcal{H}_1$. Conversely, a Bayes factor less than one suggests that $\mathcal{H}_1$ is more in agreement with the observed data. A value approximately equal to one implies that the data do not distinctly favor any model, indicating similar levels of empirical support for $\mathcal{H}_0$ and $\mathcal{H}_1$.
To interpret the Bayes factor effectively, various categorizations have been proposed. One of the most notable was outlined by \citet{jeffreys1961theory}, and is shown in Table \ref{table:Jeffreys scale}.
Notably, the thresholds in Table \ref{table:Jeffreys scale} are defined for 
$\mathrm{BF}_{01} > 1$, (i.e., when the evidence favors $\mathcal{H}_0$). 
If $\mathrm{BF}_{01} < 1$, the same categories apply to the reciprocal $\mathrm{BF}_{10} = 1 / \mathrm{BF}_{01}$, 
which indicates evidence in favor of $\mathcal{H}_1$.
\begin{table}[h!]
\centering
\caption{Scale of evidence proposed by \citet{jeffreys1961theory}. }\label{table:Jeffreys scale}
\begin{tabular}{ccc}
\toprule
\(\mathrm{BF}_{01}\) & \(\log_{10}(\mathrm{BF}_{01})\) & Evidence for $\mathcal{H}_0$ \\\hline
\(1\) to \(3.2\) & \(0\) to \(0.5\) & Barely worth mentioning \\
\(3.2\) to \(10\) & \(0.5\) to \(1\) & Substantial evidence\\
\(10\) to \(31.6\) & \(1\) to \(1.5\) & Strong evidence\\
\(31.6\) to \(100\) & \(1.5\) to \(2\) & Very strong evidence\\
\(> 100\) & \(> 2\) & Decisive evidence\\
\bottomrule
\end{tabular}
\end{table}

\subsection{Hypothesis testing for the mixture weight $\omega$} \label{HT omega}
To determine how closely the replication aligns with the original study, we may perform hypothesis testing on the mixture weight parameter $\omega$. A key goal is testing whether the original and replication studies are consistent with each other, formulated as the hypothesis $\mathcal{H}_c: \omega = 1$. This hypothesis may be tested against the alternative hypothesis that suggests the data from the studies should be entirely disregarded, indicated as $\mathcal{H}_d: \omega = 0$.

Notably, the point hypothesis $\mathcal{H}_d: \omega = 0$ avoids leading to an improper mixture prior, providing a clear methodological advantage over the power prior approach \citep{pawel2023power}. Consequently, the Bayes factor derived in this context does not encounter problematic issues related to the dependence on the ratio of the two arbitrary constants, since it is based on the ratio of two well-defined marginal likelihoods
\begin{equation} \label{eqn:7}
    \begin{aligned}
        \mathrm{BF}_{\text{dc}}(\hat{\theta}_r \mid \mathcal{H}_d: \omega = 0) 
        &= \dfrac{f\{\hat{\theta}_r \mid \mathcal{H}_d: \theta \mid \omega \sim \omega \mathrm{N}(\hat{\theta}_o,\sigma^2_o) + (1-\omega) \mathrm{N}(\mu,\tau^2), \omega = 0\}}{f\{\hat{\theta}_r \mid \mathcal{H}_c: \theta \mid \omega \sim \omega \mathrm{N}(\hat{\theta}_o,\sigma^2_o) + (1-\omega) \mathrm{N}(\mu,\tau^2), \omega = 1\}} \\
        &= \dfrac{\mathrm{N}(\hat{\theta}_r \mid \mu, \sigma^2_r+ \tau^2)}{\mathrm{N}(\hat{\theta}_r \mid \hat{\theta}_o, \sigma^2_r+ \sigma^2_o)}.
    \end{aligned}
\end{equation}

A more flexible hypothesis to consider is that the data exhibit a certain level of compatibility or disagreement.
A suitable hypothesis is defined by the prior class $\mathcal{H}_d: \omega \sim \mathrm{Beta}(1, \nu)$, where $\nu > 1$. In this class of distributions, the density is maximized at $\omega = 0$ and decreases consistently from there. This encodes a hypothesis where the importance of the original data is systematically reduced. The degree of this reduction is dictated by the parameter $\nu$. In the asymptotic case where $\nu \rightarrow \infty$, the hypothesis simplifies to $\mathcal{H}_d: \omega = 0$, implying a complete discounting of the original data. Consequently, the Bayes factor is 
\begin{equation*}
    \begin{aligned}
        \mathrm{BF}_{\text{dc}}\{\hat{\theta}_r \mid \mathcal{H}_d: \omega \sim \mathrm{Beta}(1, \nu)\} 
        &= \dfrac{f\{\hat{\theta}_r \mid \mathcal{H}_d: \theta \mid \omega \sim \omega \mathrm{N}(\hat{\theta}_o,\sigma^2_o) + (1-\omega) \mathrm{N}(\mu,\tau^2), \omega \sim \mathrm{Beta}(1, \nu) \}}{f\{\hat{\theta}_r \mid \mathcal{H}_c: \theta \mid \omega \sim \omega \mathrm{N}(\hat{\theta}_o,\sigma^2_o) + (1-\omega) \mathrm{N}(\mu,\tau^2), \omega = 1\}} \\
        &= \dfrac{\left(\dfrac{1}{1+\nu}\right)  \left\{\mathrm{N}(\hat{\theta}_r \mid \hat{\theta}_o, \sigma^2_r +\sigma^2_o)-\mathrm{N}(\hat{\theta}_r \mid \mu, \sigma^2_r + \tau^2)\right\} + \mathrm{N}(\hat{\theta}_r \mid \mu, \sigma^2_r+ \tau^2)}{\mathrm{N}(\hat{\theta}_r \mid \hat{\theta}_o, \sigma^2_r+ \sigma^2_o)}.   
    \end{aligned}
\end{equation*}

\subsection{Hypothesis testing for the effect size $\theta$} \label{HT theta}
In the assessment of hypotheses regarding the magnitude of the effect size $\theta$, the analysis typically involves a comparative evaluation between the null hypothesis, $\mathcal{H}_0: \theta = 0$, which posits absence of the effect, and the alternative hypothesis, $\mathcal{H}_1: \theta \neq 0$, suggesting the presence of an effect. The null hypothesis $\mathcal{H}_0$ represents a singular value within the possible range of $\theta$ values, while the alternative hypothesis $\mathcal{H}_1$ requires a prior specification for both $\theta$ and $\omega$.

To address this, the use of a mixture prior as in equation (\ref{eqn:1}) is proposed. Specifically, the first mixture prior component is based on the empirical data from the original study $\hat{\theta}_o$ while the second component is designed to have the same amount of information content equivalent to a single observation. This approach is complemented by the specification of a suitable Beta prior for the weight parameter $\omega$. Consequently, the Bayes factor is 
\begin{equation*}
    \begin{aligned}
        \mathrm{BF}_{01}\{\hat{\theta}_r \mid \mathcal{H}_1: \omega \sim \mathrm{Beta}(\eta,\nu)\} 
        &= \dfrac{f(\hat{\theta}_r \mid \mathcal{H}_0: \theta = 0)}{f\{\hat{\theta}_r \mid \mathcal{H}_1: \theta \mid \omega \sim \omega \mathrm{N}(\hat{\theta}_o,\sigma^2_o) + (1-\omega) \mathrm{N}(\mu,\tau^2), \omega \sim \mathrm{Beta}(\eta,\nu) \}} \\
        &= \dfrac{\mathrm{N}(\hat{\theta}_r \mid 0, \sigma^2_r)}{\left(\dfrac{\eta}{\eta+\nu}\right)  \left\{\mathrm{N}(\hat{\theta}_r \mid \hat{\theta}_o, \sigma^2_r +\sigma^2_o)-\mathrm{N}(\hat{\theta}_r \mid \mu, \sigma^2_r + \tau^2)\right\} + \mathrm{N}(\hat{\theta}_r \mid \mu, \sigma^2_r+ \tau^2)}.
    \end{aligned}
\end{equation*}

It is important to emphasize that,  as parallel discussed in \citet{pawel2023power} for the power parameter in the power prior approach, assigning a point mass to the weight parameter $\omega = 1$ leads to the Bayes factor contrasting a point null hypothesis to the posterior distribution of the effect size based on the original data that is the \textit{replication Bayes factor} under normality \citep{verhagen2014bayesian, ly2019replication,pawel_held_2022}. In detail, it is
\begin{equation} \label{eqn:rep_BF}
    \begin{aligned}
        \mathrm{BF}_{01}\{\hat{\theta}_r \mid \mathcal{H}_1: \omega = 1\} 
        &= \dfrac{f(\hat{\theta}_r \mid \mathcal{H}_0: \theta = 0)}{f\{\hat{\theta}_r \mid \mathcal{H}_1: \theta \mid \omega \sim \omega \mathrm{N}(\hat{\theta}_o,\sigma^2_o) + (1-\omega) \mathrm{N}(\mu,\tau^2), \omega = 1\}} \\
        &= \dfrac{\mathrm{N}(\hat{\theta}_r \mid 0, \sigma^2_r)}{\mathrm{N}(\hat{\theta}_r \mid \hat{\theta}_o, \sigma^2_r + \sigma^2_o)}.
    \end{aligned}
\end{equation}
Similar to the power prior formulation, the mixture prior version of the replication Bayes factor represents a generalization of the standard replication Bayes factor that provides a flexible and controlled approach for combining original and replication data.

\subsection{Posterior distribution and Bayes factor asymptotics}
To explore the proposed mixture model further, a key focus is on examining the asymptotic characteristics of the marginal posterior distribution and the Bayes factor for the weight parameter. Specifically, let us consider the Bayes factor contrasting $\mathcal{H}_d \colon \theta \sim \mathrm{N}(\mu, \tau^2)$ to $\mathcal{H}_c \colon \theta \sim \mathrm{N}(\hat{\theta}_o, \sigma^2)$ for the replication data $\hat{\theta}_r \mid \theta \sim \mathrm{N}(\theta, \sigma^2_r)$ as in (\ref{eqn:7}).
Subsequently, the marginal posterior distribution in (\ref{eqn:6}) can be expressed in terms of the Bayes factor
\begin{equation}
\begin{aligned}
    \pi(\omega \mid \hat{\theta}_r, \hat{\theta}_o, \eta, \nu) 
    &= \frac{\pi(\omega) \left\{\omega \mathrm{N}(\hat{\theta}_r \mid \hat{\theta}_o, \sigma^2_r + \sigma^2_o) + (1 - \omega) \mathrm{N}(\hat{\theta}_r \mid \mu, \sigma^2_r + \tau^2)\right\}}{\left(\dfrac{\eta}{\eta+\nu}\right)  \left\{ \mathrm{N}(\hat{\theta}_r \mid \hat{\theta}_o, \sigma^2_r + \sigma^2_o) - \mathrm{N}(\hat{\theta}_r \mid \mu, \sigma^2_r + \tau^2)\right\} + \mathrm{N}(\hat{\theta}_r \mid \mu, \sigma^2_r + \tau^2)} \\
    &= \frac{\pi(\omega)\left\{\omega  + (1 - \omega)\mathrm{BF}_{\text{dc}}(\hat{\theta}_r)\right\}}{\left(\dfrac{\eta}{\eta+\nu}\right)\left\{1 - \mathrm{BF}_{\text{dc}}(\hat{\theta}_r)\right\} + \mathrm{BF}_{\text{dc}}(\hat{\theta}_r)}. 
\end{aligned}
\label{eqn:8}
\end{equation}
We investigate the behavior of the limiting marginal posterior distribution in (\ref{eqn:8}) when the Bayes factor tends to zero and when it tends towards infinity, respectively. In these cases, we have that
\begin{equation*}
   \begin{aligned}
    \lim_{\mathrm{BF}_{\text{dc}}(\hat{\theta}_r) \downarrow 0} \pi(\omega \mid \hat{\theta}_r, \hat{\theta}_o, \eta, \nu)
    &= \frac{\pi(\omega)}{\mathbb{E}_{\pi(\omega)}(\omega)}  \omega = \mathrm{Beta}(\omega \mid \eta + 1, \nu),\\
    \lim_{\mathrm{BF}_{\text{dc}}(\hat{\theta}_r) \uparrow +\infty} \pi(\omega \mid \hat{\theta}_r, \hat{\theta}_o, \eta, \nu)
    &= \frac{\pi(\omega) }{1 - \mathbb{E}_{\pi(\omega)}(\omega)} (1 - \omega) = \mathrm{Beta}(\omega \mid \eta, \nu + 1). 
   \end{aligned}
\end{equation*}
This means that even when we find overwhelming evidence in favor of $\mathcal{H}_d$ or $\mathcal{H}_c$, the posterior distribution is only slightly changed from the prior (i.e., ``updated by one observation'' from the prior). For example, for a flat prior with  $\nu = \eta = 1$, the limiting posteriors are given by the $\mathrm{Beta}(2, 1)$ and $\mathrm{Beta}(1, 2)$ distributions, respectively, which correspond to densities that are linearly increasing (decreasing) from 0 (2) to 2 (0). We can see from Figure~\ref{plot_marg_post_joint} that the marginal posteriors for the second, third, and pooled ``Moral credentialing'' replications are not too different from these two asymptotic distributions. 

\begin{figure*}[htb!]
    \centering
        \includegraphics[width=\textwidth, height=8cm]{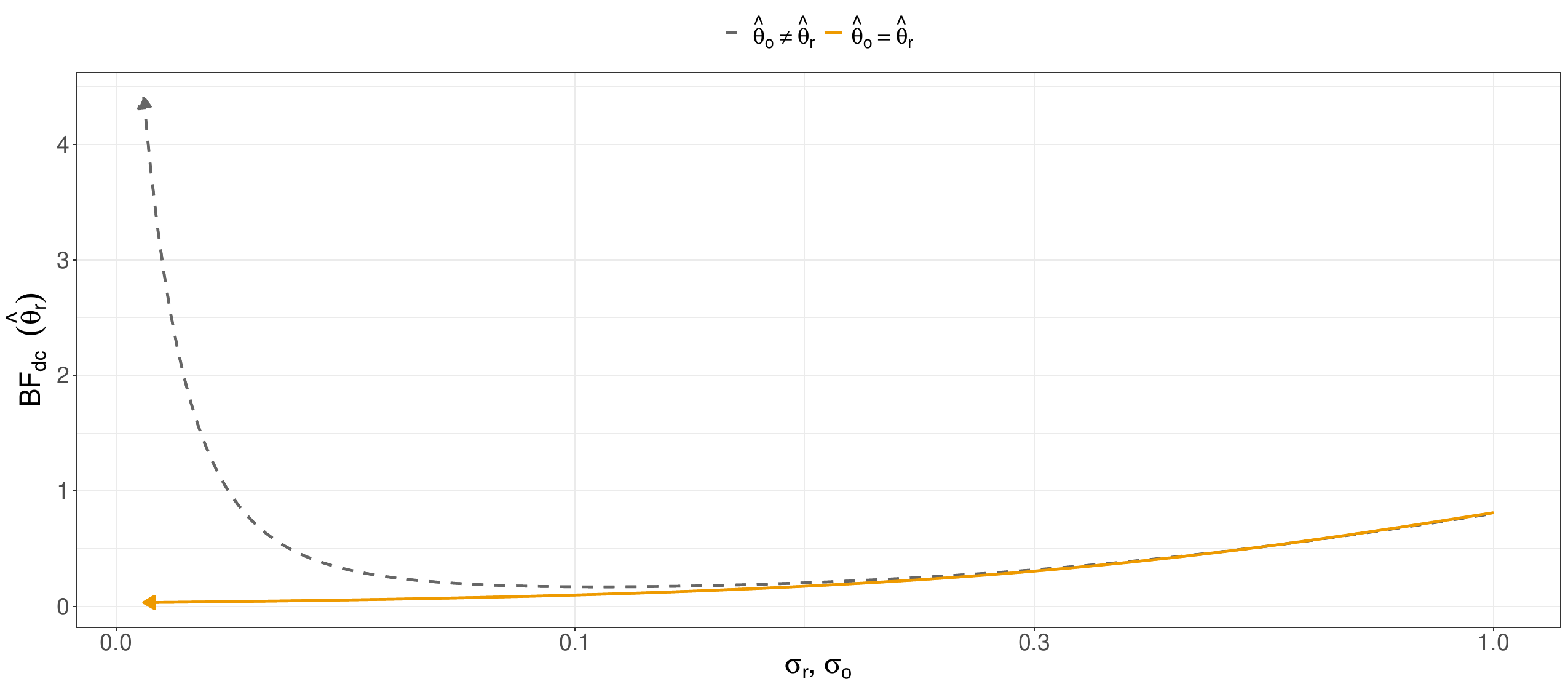}
        \caption{Asymptotic behavior of the Bayes factor for the weight parameter $\mathrm{BF}_{\text{dc}}(\hat{\theta}_r)$  as both original and replication standard errors tend to zero.}
        \label{bf_asymp}
\end{figure*}

While the previous calculations assumed that the Bayes factor can go to infinity or zero, thereby overwhelmingly favoring one of the contrasted models, it is unclear whether this is even possible. We therefore  now  aim to assess the consistency property of the Bayes factor \citep{kass_raftery_1995, chib2016bayes} so that when one model truly generates the data, the Bayes factor will increasingly favor that model as sample size grows. We explore the effects on the Bayes factor (\ref{eqn:7}) when the standard error of the replication study $\sigma_r$ becomes arbitrarily small. This could occur due to an increase in the sample size, which is typically inversely related to the squared standard error. The limiting Bayes factor as the replication standard error $\sigma_r$ approaches zero is
\begin{align*}
    \lim_{\sigma^2_r \downarrow 0} \mathrm{BF}_{\text{dc}}(\hat{\theta}_r)
    &= \frac{\mathrm{N}(\hat{\theta}_r \mid \mu, \tau^2)}{\mathrm{N}(\hat{\theta}_r \mid \hat{\theta}_o, \sigma^2_o)}.
\end{align*}
Consequently, for finite $\tau^2$ and $\sigma^2_o$, the Bayes factor is bounded and cannot converge to either zero or $+\infty$. However, if the original standard error $\sigma_o$ also approaches zero, the Bayes factor in (\ref{eqn:7}) behaves differently. In this case, the Bayes factor approaches
\begin{align*}
    \lim_{\sigma^2_r, \sigma^2_o \downarrow 0} \mathrm{BF}_{\text{dc}}(\hat{\theta}_r)
    &= \frac{\mathrm{N}(\hat{\theta}_r \mid \mu, \tau^2)}{\delta_{\hat{\theta}_o}(\hat{\theta}_r)},
\end{align*}
where $\delta_{\hat{\theta}_o}(\cdot)$ represents the Dirac delta function. When the standard errors from both original and replication go to zero, the Bayes factor thus shows the correct asymptotic behavior, converging to zero when their effect sizes are the same while converging to infinity when they are not,
as illustrated in Figure \ref{bf_asymp}.  

\subsection{Hypothesis testing for the ``Moral credentialing'' experiment} \label{HT Labels}

We will now illustrate the results of the proposed hypothesis tests in Sections \ref{HT omega} and \ref{HT theta} using the data obtained from the ``Moral credentialing'' experiment, as described in Section \ref{Running example}.
Specifically,  a comprehensive analysis is performed to understand the behaviour of these parameters in the replication study.

To evaluate the agreement between the initial study and subsequent replications, Table \ref{tab:bf_omega} shows the results of the hypothesis tests concerning the mixture weight parameter $\omega$. The fourth column shows the Bayes factor contrasting two point hypotheses: $\mathcal{H}_d: \omega = 0$ and $\mathcal{H}_c: \omega = 1$. This analysis reveals substantial and strong evidence for $\mathcal{H}_c$ in the first, second, and pooled replication scenarios, respectively. Conversely, the third replication study shows strong evidence favoring $\mathcal{H}_d$. 
While the Bayes factors based on the $\mathrm{Beta}(1, 2)$ prior under $\mathcal{H}_d$ (fifth column) still point in the same direction, the extent of evidence is lower than for the point hypothesis.
\begin{table*}[h!]
    \centering
    \caption{``Moral credentialing'' experiment. Hypothesis tests for the mixture weight $\omega$.}
\begin{tabular}{@{}ccccc@{}}
  \toprule Replication & $\hat{\theta}_r$ & $\sigma_r$ & $\mathrm{BF}_{\text{dc}}(\hat{\theta}_r \mid \mathcal{H}_d : \omega = 0)$ & $\mathrm{BF}_{\text{dc}}\{\hat{\theta}_r \mid \mathcal{H}_d : \omega \sim \mathrm{Beta}(1, 2)\}$ \\ 
  \midrule
  1 & 0.29 & 0.11 & 1/9.3 & 1/2.5  \\ 
    2 & 0.25 & 0.09 & 1/12 & 1/2.6  \\ 
    3 & -0.18 & 0.11 & 9.9 & 7  \\ 
    Pooled & 0.14 & 0.06 & 1/13 & 1/2.6 \\ 
   \bottomrule
\end{tabular}
    \label{tab:bf_omega}
\end{table*}

Table \ref{tab:bf_theta} presents the outcomes of our hypothesis tests regarding the effect size parameter $\theta$. Specifically, the fourth column shows the Bayes factors contrasting the null hypothesis ($\mathcal{H}_0: \theta = 0$) to the alternative hypothesis ($\mathcal{H}_1: \theta \neq 0$), with a $\mathrm{Beta}(1,1)$ prior for the weight parameter $\omega$ under $\mathcal{H}_1$. The results suggest that there is moderate to strong evidence for a non-zero effect in the first replication, second, and pooled replications. Conversely, the Bayes factor $\mathrm{BF}_{01}\{\hat{\theta}_r \mid \mathcal{H}_{1} : \omega \sim \mathrm{Beta}(1, 1)\}$ indicates strong evidence in favor of $\mathcal{H}_0$ for the third replication.  In addition, the evidence from the replication Bayes factor under normality, $\mathrm{BF}_{01}(\hat{\theta}_r \mid \mathcal{H}_{1} : \omega = 1)$ leads to the same qualitative conclusions thus corroborating the previous findings.
\begin{table*}[h!]
    \centering
\caption{``Moral credentialing'' experiment. Hypothesis tests for the effect size $\theta$.}
\begin{tabular}{@{}ccccc@{}}
  \toprule 
 Replication & $\hat{\theta}_r$ & $\sigma_r$  & $\mathrm{BF}_{01}(\hat{\theta}_r \mid \mathcal{H}_{1} : \omega = 1)$ & $\mathrm{BF}_{01}\{\hat{\theta}_r \mid \mathcal{H}_{1} : \omega \sim \mathrm{Beta}(1, 1)\}$ \\ 
  \midrule
  1 & 0.29 & 0.11 & 1/21 & 1/12 \\ 
    2 & 0.25 & 0.09 & 1/38 & 1/21 \\ 
    3 & -0.18 & 0.11 & 34 & 6.2 \\ 
     Pooled & 0.14 & 0.06 & 1/8.1 & 1/4.3 \\
   \bottomrule
\end{tabular}
    \label{tab:bf_theta}
\end{table*}

In summary, the findings from our analysis indicate that among the three replications, the first and second replications align with the original study's results and also offer evidence for a non-zero effect. Conversely, the third replication presents evidence for the absence of an effect and does not align with the findings of the original study. However, when pooled, the replications align with the original study's findings and provide evidence for a non-zero effect, indicating that the replication effort was successful overall.

\begin{figure*}[htb!]
    \centering
        \includegraphics[width=\textwidth, height=8cm]{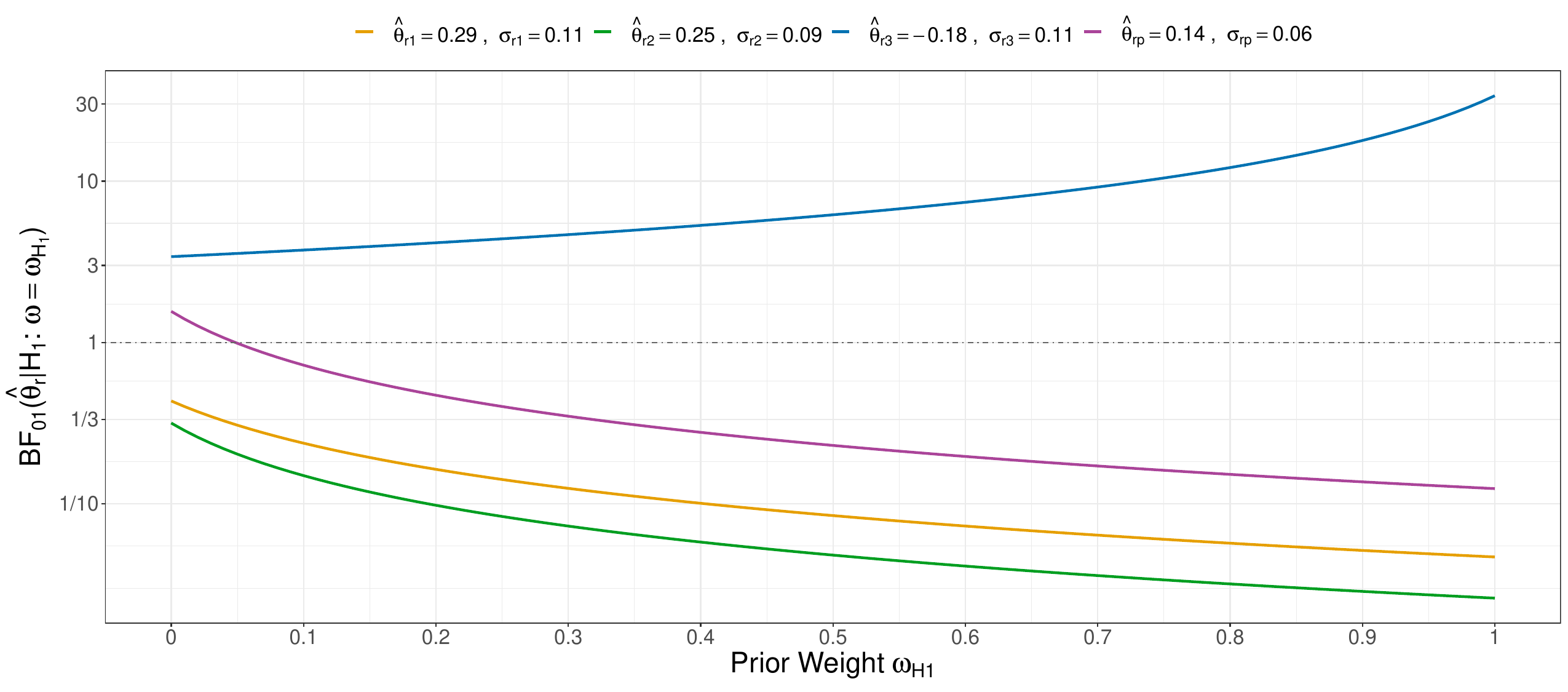}
        \caption{``Moral credentialing'' experiment. Tipping-point analysis for the Bayes factor $\mathrm{BF}_{01}\{\hat{\theta}_r \mid \mathcal{H}_1: \omega = \omega_{\mathcal{H}_1}\}$}
        \label{tipping_point_BF_theta}
\end{figure*}

Furthermore, similar to the approach described in Section \ref{Fixed weight parameter}, we perform a tipping-point analysis using Bayes factors instead of credible intervals. Specifically, we treat the Bayes factor as a function of the fixed mixing weight $\omega$ under the alternative hypothesis and investigate whether there is a ``tipping'' value at which the Bayes factor equals $1$, meaning equal evidence for both hypotheses. 
For the effect‐size $\theta$, we use a Bayes factor similar to $\eqref{eqn:rep_BF}$ that contrasts the point null hypothesis $\mathcal{H}_0: \theta = 0$ against the posterior distribution of $\theta$ with a specified fixed weight parameter $\omega_{\mathcal{H}_1}$. As shown in Figure \ref{tipping_point_BF_theta}, the first and second replications yield consistent conclusions across the entire range of $\omega_{\mathcal{H}_1}$, demonstrating increasing evidence in favor of the alternative hypothesis $\mathcal{H}_1: \omega = \omega_{\mathcal{H}_1}$, which implies the presence of an effect. In contrast, the third replication shows growing evidence for the null hypothesis of no effect, $\mathcal{H}_0: \theta = 0$, with increasing weight. For neither of the three individual replications, there is a tipping-point where the conclusions qualitatively change (i.e., where Bayes factor crosses 1). However, for the pooled replication, a minimum mixture weight $\omega_{\mathcal{H}_1} = 0.056$ is required to obtain evidence supporting the presence of an effect. For smaller weights, there is anecdotal evidence in favor of the null hypothesis. This result differs from the credible interval tipping-point analysis in Figure~\ref{hpdi_fixed_weights} where there is no tipping-point at which the credible interval excludes an effect size of zero, demonstrating the difference between Bayes factors and posterior credible interval inferences.

\section{Comparison with power priors and hierarchical models} \label{sec:comparison}

In this section, we compare our mixture prior framework with two other prominent Bayesian approaches for replication studies: Power priors \citep{pawel2023power} and hierarchical models \citep{bayarri_2002a, bayarri_2002b, Pawel_Held_2020}.  We focus on conceptual differences, practical implementation, and inferential behavior.

\subsection{Power priors}
Power priors \citep{chen_ibrahim2000} are a popular class of informative prior distributions that allow incorporating historical data into a current analysis. In its basic version, the power prior is obtained by updating an initial (often non-informative) prior $\pi_0(\theta)$ by the historical data likelihood raised to a power parameter $\alpha$, where $0\le\alpha\le1$. Under the assumption that the likelihood of the effect estimates can be approximated by a normal distribution and with an (improper) flat initial prior $\pi_0(\theta)\propto 1$ the resulting power prior is
\begin{equation}
\label{eqn:pp}
      \pi_{\mathrm{PP}}(\theta \mid \hat\theta_o, \alpha) =
  \mathrm{N}(\theta \mid \hat\theta_o, \sigma^2_o/\alpha).
\end{equation}
Here, $\alpha$ determines the extent to which historical data influence the prior distribution. Specifically, setting $\alpha$ to zero completely discounts historical information, while setting it to one fully integrates historical information into the prior.  In the formulation as in Equation \eqref{eqn:pp} the power parameter is often fixed at a reasonable value based on previous knowledge or using empirical Bayes estimates \citep{Gravestock2017, Gravestock_2019}. 

Alternatively, uncertainty about the power parameter $\alpha$ can be explicitly modeled by specifying a prior distribution for it. The most common choice is a beta distribution $\alpha \mid \eta, \nu \ \sim \mathrm{Beta} (\eta, \nu)$ leading to the normalized power prior \citep{duan2006, Neuenschwander2009} with density
\begin{equation}
\label{eqn:npp}
      \pi_{\mathrm{NPP}}(\theta \mid \hat\theta_o, \alpha) =
  \mathrm{N}(\theta \mid\hat\theta_o, \sigma^2_o/\alpha) \mathrm{Beta}(\alpha \mid \eta, \nu).
\end{equation}
A uniform prior distribution is commonly recommended as the default choice for the power parameter \citep{Ibrahim2015}.

\subsection{Hierarchical models}
Hierarchical modeling is another approach that allows the incorporation of historical data in Bayesian analyses, with meta-analysis representing one of its most common applications \citep[among others]{smith1995bayesian, sutton2000methods, spiegelhalter2004bayesian, lunn2013fully}. Within this context, two distinct perspectives on meta-analysis have emerged. The traditional perspective (also called literature synthesis meta-analysis), aims to combine results across studies to estimate an average effect. In contrast, Rubin’s response-surface perspective \citep{rubin1990new} treats meta-analysis as a modeling and extrapolation problem. The goal is to infer the “true” scientific effect -- namely, the effect that would be observed in an idealised, infinitely large, perfectly designed study -- by incorporating study-level predictors and estimating a response surface using meta-regression techniques. For further details see \citet{rubin1990new}, \citet{rubin1992meta} and  \citet{zhang2023towards}

In the replication setting, a standard formulation is a two-level normal–normal model
\begin{equation}
\begin{aligned}
\label{eqn:hier_mod}
\hat{\theta}_i \mid \theta_i &\sim \mathrm{N}(\theta_i,\;\sigma_i^2), \quad i \in \{o,r\},\\ \theta_i \mid \mu,\;\tau^2 &\sim \mathrm{N}(\mu,\;\tau^2), \quad i \in \{o,r\},
\end{aligned} 
\end{equation}
where $\mu$ represents the underlying overall effect size, while $\tau^2$ is the variance (or squared heterogeneity). 
Typically $\mu$ is given a vague prior (e.g., flat) that does not reflect a strong prior belief in the true effect. The heterogeneity parameter $\tau$ controls how similar the true effects of the original and the replication studies are expected to be,  acting as a shrinkage tool \citep{gelman2006prior}.  If $\tau$ is close to zero, the model assumes that the effect sizes are essentially the same (complete pooling); larger $\tau$ allows the two studies’ true effects $\theta_o$ and $\theta_r$ to differ more (less pooling). It is important to note that in the replication setting, the interest lies in estimating and testing the replication-specific effect size $\theta_r$ and the heterogeneity $\tau$, and less in the original-specific effect size $\theta_o$ and the overall effect size $\mu$, the latter parameters simply providing a means to integrate the original data into the analysis.

\subsection{Comparison}
We will now investigate how these approaches to assessing replication success compare with each other across the four replication scenarios from the ``Moral credentialing'' experiment. We focus on evaluating how each approach incorporates information from the original study, handles discrepancies between original and replication data, and influences the resulting posterior inference. For the power prior, we used a normalized formulation as in \eqref{eqn:npp} with a $\mathrm{Beta}(1,1)$ prior for the power parameter $\alpha$, using the R package \texttt{ppRep} \citep{pawel2023power}. For the hierarchical model, we used the R package \texttt{bayesmeta} \citep{bayesmeta} to fit a two-level normal–normal model \eqref{eqn:hier_mod} with an improper uniform prior on $\mu$ and a $\mathrm{half\text{-}normal}$ prior on $\tau$. 
Although a $\mathrm{half\text{-}normal}(0.2)$ prior has been recommended as weakly informative for meta-analyses \citep[p.468]{Roever2021}, we expect less heterogeneity in direct replication scenarios compared to the meta-analytic context. Therefore, we selected a  $\mathrm{half\text{-}normal}(0.1)$ prior for $\tau$ to reflect the reduced variability.

\begin{figure*}[htb!]
    \centering
        \includegraphics[width=\textwidth, height=8cm]{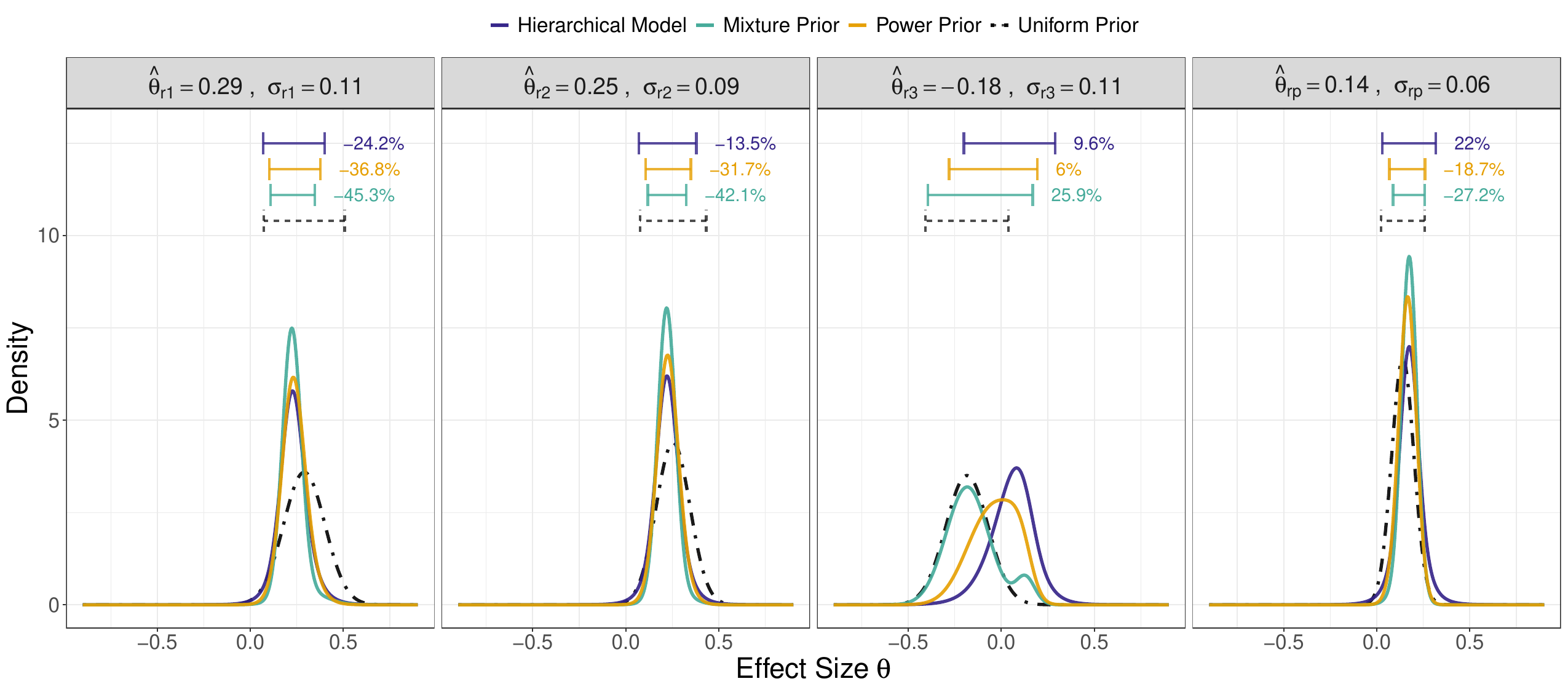}
        \caption{``Moral credentialing'' experiment. Marginal posterior distributions and $95\%$ highest posterior density interval (HPDI) of the effect size $\theta$ using different approaches. The dashed lines represent the posterior density of the effect size $\theta$, derived exclusively from the replication data, without considering the original data, and assuming a uniform prior for the effect size $\pi(\theta)\propto 1$. The  numbers represent the percentage difference in HPDI length relative to the replication‐only analysis (positive values indicate HPDIs longer than the replication‐only analysis, negative values indicate HPDIs shorter).}
        \label{comp_methods}
\end{figure*}

 Figure \ref{comp_methods} shows the marginal posterior distributions for the effect size $\theta$ for each replication. The plot is enriched by contrasting it with the posterior distribution of $\theta$ based solely on the replication data, represented as a dashed line. For the first, second, and pooled replications (where the original study’s result aligns well with the replication data), the mixture prior, the power prior, and the hierarchical model borrow substantial information from the original study. This leads to more concentrated posteriors for $\theta$, with substantially shorter $95\%$ HPDIs than those obtained using only the replication data. In particular, the mixture prior yields the sharpest and narrowest posterior distribution among the methods, reducing the HPDI length by $45.3\%$, $42.1\%$, and $27.2\%$ for the first, second, and pooled replications, respectively, relative to the replication‐only analysis. For the third replication scenario, where the replication effect is in conflict with the original finding, the mixture and power prior approaches discount the original data almost entirely. Consequently, the posterior remains centered near the replication’s negative effect estimate rather than being pulled toward the original study’s positive estimate. Furthermore, the resulting $95\%$ HPDI are notably wider than those of the replication-only posterior. The hierarchical model similarly detects the large discrepancy between the original and replication results by inflating the heterogeneity parameter $\tau$. As a result, the hierarchical model’s posterior for $\theta_r$ becomes more diffuse -- even broader than the replication only posterior -- and shifts toward zero.

Both the mixture prior and power prior approaches can be viewed as special cases of robust Bayesian updating, where an extra parameter is introduced to control the influence of historical (original) data. In addition, \citet{pawel2023power} demonstrated a connection between power priors and hierarchical models: Under the meta-analytic framework, the power parameter $\alpha$ corresponds to the relative heterogeneity measure $I^2 = \tau^2/(\tau^2 + \sigma^2_o)$ \citep{higgins2002quantifying} via $\alpha = (1 - I^2)/(1 + I^2)$. From an analyst perspective, the mixture prior is more intuitive  with respect to the power prior approach due to the weight parameter, which provides a straightforward measure to quantify how much information is discounted. In our example, the mixture prior appears to be more efficient in assessing how much original and replicated data should be pooled, producing the shortest HPDI when the data agree and the widest when they disagree. From a computational perspective, the mixture prior is more convenient because it yields a closed-form posterior under the meta-analytic framework. Conversely, the normalized power prior requires a normalizing constant calculation that is  computationally intensive \citep{Lesaffre2024}. Finally, while the hierarchical model is a very general and comprehensive approach, it requires specifying priors for both the overall effect size $\mu$ and the heterogeneity variance $\tau^2$, adding an additional layer of modeling complexity.

\section{Discussion} \label{Discussion}
In this paper, we introduced a novel Bayesian method for analyzing data from replication studies.
By using a mixture prior that mixes the posterior based on the original study with a non-informative prior, our method addresses the issue of potential conflict between original and replication study, as in such cases the information from the original study can be discounted. A crucial element is the mixture weight parameter $\omega$. 
We explored two distinct strategies for setting this weight parameter. The first strategy involves fixing the weight to a specific value, for example, on the basis of expert knowledge or an empirical Bayes estimate. The sensitivity of this choice may then be assessed with a reverse-Bayes tipping-point analysis \citep{Best2021, held_reverse_bayes}.
The second strategy introduces a level of uncertainty by assigning a prior distribution to the mixture weight parameter. We then showed that the prior on the weight strategy is equivalent to the fixed weight strategy using the expected value of the weight's prior as fixed weight. However, the uncertain weight strategy also provides data analysts with a posterior distribution of the weight, which can be used for quantitatively assessing the degree of study compatibility, yet the extent to which this posterior can be updated from the prior was also shown to be limited. Importantly, both strategies yield the same results for the effect size when the fixed weight is equal to the expectation of the prior distribution. The only difference lies in the additional posterior distribution for the weight parameter. 

Scientists should choose between these two strategies based on the characteristics of their study. Fixed weights can be more straightforward and are based on prior knowledge. 
They are suitable in situations where there is a reasonable confidence about the degree of agreement between the original and replication studies. A tipping-point analysis can additionally help to assess how robust the analysis is to the choice of the weight. On the other hand, the random weight approach provides an additional posterior distribution for the weight parameter, showing the uncertainty related to this parameter.

We also presented Bayesian hypothesis tests for assessing the magnitude of the effect size $\theta$ and to determine how closely the replications align with the original study. We analyzed the asymptotic behavior of the marginal posterior distribution for the weight parameter 
when the Bayes factor tends to zero or towards infinity. Moreover, we examined how the Bayes factor related to the effect size behaves as the replication study's standard error $\sigma_r$ tends to zero. Our findings reveal that the Bayes factor contrasting $\mathcal{H}_d \colon \theta \sim \mathrm{N}(\mu, \tau^2)$ to $\mathcal{H}_c \colon \theta \sim \mathrm{N}(\hat{\theta}_o, \sigma^2_o)$, for finite $\tau^2$ and $\sigma^2_o$, is inconsistent. However, when the original study's standard error $\sigma_o$ also approaches zero, the behavior of the Bayes factor 
changes, leading to correct asymptotic behavior and consistency.

The mixture prior approach we developed presents some similarities with two well-established methods in the replication setting -- power priors \citep{pawel2023power} and hierarchical models \citep{bayarri_2002a,bayarri_2002b,Pawel_Held_2020}. All three approaches exhibit similar strengths in assessing differences between original and replication studies, providing valuable inferences that complement each other. Analogously to the heterogeneity variance and the power parameter, the mixture weight $\omega$ controls the degree of compatibility between the original and replication studies. Nevertheless,  our approach offers some advantages. First, we believe that the mixture weight parameter $\omega$ provides a more straightforward and intuitive discounting measure compared to the power parameter used in the power prior approach, making this approach more accessible for analysts. Second, the inherent structure of the proposed mixture prior provides methodological advantages over power priors, particularly concerning hypothesis testing for determining how much original and replicated data should be pooled. Specifically, our approach avoids the risk of generating an improper prior when testing consistency between original and replication studies. Third, the mixture prior also has some computational advantages. Notably, the calculation of the marginal likelihood in the random weight scenario is similar to that in the fixed weight scenario, with the only difference being the replacement of the fixed weight with the expected weight over the prior, which is computationally advantageous. This is particularly evident when compared to the computationally-prohibitive normalizing constant of the normalized power prior \citep{Lesaffre2024}. Finally, when multiple original studies are involved, our mixture prior approach may facilitate their inclusion in the analysis. Specifically, this can be achieved by using two or more informative components derived from the original studies, along with a non-informative component.

Our method relies on the widely-used meta-analytic assumption that the distribution of effect estimates can be accurately approximated by a normal distribution with known variance, making it adaptable to a broad range of effect sizes from various data models across different research fields. However,
this assumption becomes too strong in presence of small sample sizes and/or extreme effect size values at the boundary of the parameter space (e.g., very small or large probabilities). Future research could thus adapt our approach to specific data models (e.g., binomial or Student's t distribution), especially in the presence of small sample sizes. 

In this paper, we analyze replications both individually by directly comparing each one with the original study, and simultaneously by pooling them into a unique replication without assuming heterogeneity among the replications. An alternative pooling approach would be to assume a hierarchical model for the replication effect sizes that incorporates potential between-replication heterogeneity with a heterogeneity variance parameter.  Similarly, the mixture prior could be combined with a hierarchical prior in case there are multiple original studies for a replication study.  However, in both cases, it remains unclear how to specify the heterogeneity parameter or a prior distribution for it. Consequently, an opportunity for future research could be to explore methods to specify a fixed value for this additional parameter or to elicit a prior distribution for it.


\section*{Software and Data Availability}
All analyses were conducted in the R programming language version $4.4.3$ \citep{R_software_2025}. The code and data to reproduce this manuscript are openly available at \url{https://github.com/RoMaD-96/MixRep}. We provide an R package \texttt{repmix} for analysis of replication studies using the mixture prior framework. The package is currently available on GitHub and can be installed by running \texttt{remotes::install\_github(repo = "RoMaD-96/repmix")} (requiring the \texttt{remotes} package available on CRAN). We plan to release the package on CRAN in the future.

\bibliographystyle{apalike}
\bibliography{main}

\end{document}